\documentclass[preprint,12pt]{elsarticle}

\usepackage{amssymb}
\usepackage[utf8]{inputenc}
\usepackage[version-1-compatibility]{siunitx}

\journal{Nuclear Physics B}

\begin{document}

\begin{frontmatter}



\title{Molecular Dynamics Simulation of Apolipoprotein E3 Lipid Nanodiscs}

\author[inst1]{Patrick Allen}
\author[inst1]{Adam C. Smith}
\author[inst1]{Vernon Benedicto}
\author[inst1]{Abbas Abdulhassan}
\author[inst1]{Vasanthy Narayanaswami}
\author[inst1]{Enrico Tapavicza\corref{cor1}}
\ead{enrico.tapavicza@csulb.edu}
\cortext[cor1]{enrico.tapavicza@csulb.edu}

\affiliation[inst1]{organization={Department of Chemistry and Biochemistry, California State University, Long Beach},
            addressline={1250 Bellflower Boulevard}, 
            city={Long Beach},
            postcode={90840}, 
            state={State One},
            country={USA}}

\begin{abstract}
Nanodiscs are binary discoidal complexes of a phospholipid bilayer circumscribed by belt-like helical scaffold proteins. Using coarse-grained and all-atom molecular dynamics simulations, we explore the stability, size, and structure of nanodiscs formed between the N-terminal domain of apolipoprotein E3 (apoE3-NT) and variable number of 1,2-dimyristoyl-sn-glycero-3-phosphocholine (DMPC) molecules. We study both parallel and antiparallel double-belt configurations, consisting of four proteins per nanodisc. Our simulations predict nanodiscs containing between 240 and 420 DMPC molecules to be stable. The antiparallel configurations exhibit an average of 1.6 times more amino acid interactions between protein chains and 2 times more ionic contacts, compared to the parallel configuration. With one exception, DMPC order parameters are consistently larger in the antiparallel configuration than in the parallel one. In most cases, the root mean square deviation of the positions of the protein backbone atoms is smaller in the antiparallel configuration. We further report nanodisc size, thickness, radius of gyration, and solvent accessible surface area. Combining all investigated parameters, we hypothesize the antiparallel protein configuration leading to more stable and more rigid nanodiscs than the parallel one.
\end{abstract}

\begin{graphicalabstract}
\includegraphics[scale=0.2]{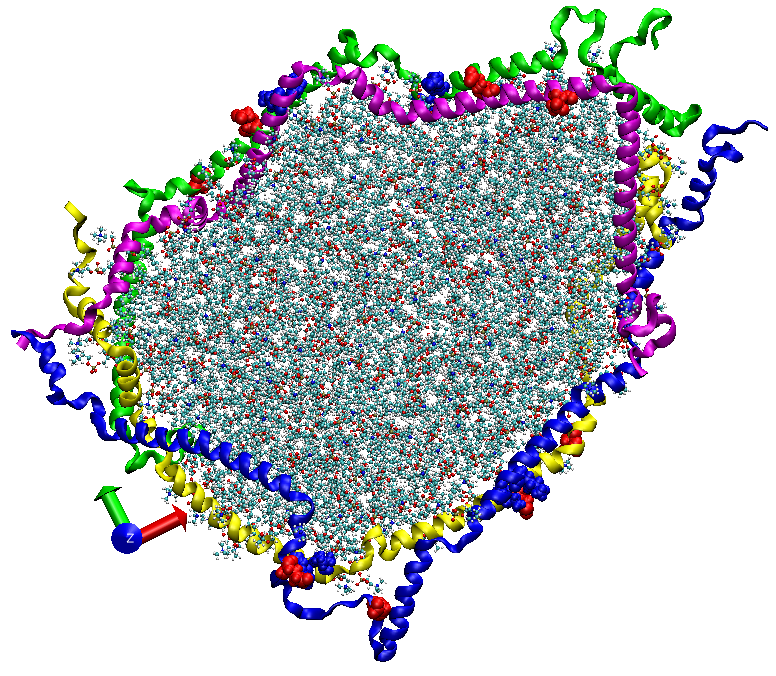}
\end{graphicalabstract}

\begin{highlights}
\item Nanodisc consisting of 4 apoE3-NT protein chains and 240 and 420 DMPC are predicted to be stable.
\item Predicted nanodisc structures exhibit diameters in accordance with experimental measurements.
\item Nanodisc with antiparallel protein configuration exhibit more ionic protein-protein contacts.
\item Most nanodisc with antiparallel protein configuration exhibit larger DMPC order parameters.
\item Most nanodisc with antiparallel protein configuration exhibit lower protein backbone RMSD.
\item Overall, we predict nanodiscs with antiparallel protein configuration to be more stable and more rigid.
\end{highlights}

\begin{keyword}
lipid nanodisc \sep apoliprotein \sep apolipoprotein E3 \sep molecular dynamics 
\PACS 0000 \sep 1111
\MSC 0000 \sep 1111
\end{keyword}

\end{frontmatter}


\section*{Introduction}
Apolipoprotein E (apoE) is a major lipid transport protein that plays a critical role in maintaining cholesterol homeostasis in the plasma and the central nervous system \cite{weisgraber1994apolipoprotein}. It belongs to the superfamily of exchangeable apolipoproteins that bear a remarkable ability to exist in both lipid-free and lipid-associated states. They have the ability to bind to and solubilize lipids to form soluble complexes called lipoproteins which are roughly spherical complexes comprising a monolayer of amphipathic lipids and proteins surrounding a core of hydrophobic lipids \cite{ference2020lipids}. The simplest lipoprotein is composed of a bilayer of phospholipids surrounded by the apolipoprotein resembling a disc termed nanodisc \cite{denisov2017nanodiscs}. They resemble the nascent high density lipoprotein (HDL) found in vivo and can be facilely assembled on the bench yielding reconstituted HDL (rHDL).

ApoE is a $\approx$34 kDa glycosylated protein composed of a series of amphipathic helices that are folded to form two domains: an N-terminal (NT) domain and a C-terminal (CT) domain. In the lipid-free state, the NT domain (residues 1-191) is composed of a four-helix bundle with the four helices (H1, H2, H3 and H4) arranged in an up-and-down manner. It is connected to the CT domain (residues 201-299) by a linker loop region (residues 192-200) \cite{wilson1991three,chetty2017helical}. In the lipid-associated state, the NT domain possesses the ability to serve as a high-affinity ligand for the low-density lipoprotein (LDL) receptor (LDLr) family of proteins by virtue of a constellation of basic residues in H4 of the 4-helix bundle and neighboring sites. In humans, apoE is polymorphic with three major alleles giving rise to apoE2, apoE3 and apoE4 isoforms, that differ from each other in one of two locations in the NT domain: position 112 or 158. ApoE2 bears a Cys at both and apoE4 bears an Arg at both, while apoE3 has a Cys at 112 and an Arg at 158. These amino acid variations result in different lipid binding abilities \cite{saito2003effects} and physiological behavior \cite{grootendorst2005human}. ApoE2 is a risk factor for cardiovascular disease, apoE4 is a risk factor for cardiovascular and Alzheimer's disease, while apoE3 is generally considered to be cardioprotective \cite{mahley2016apolipoprotein}.
Lipid-association of apoE confers the ability to recognize and interact with the LDLr. Interestingly, the isolated NT domain of apo-E (apoE-NT) retains its ability to bind the LDLr in the lipid bound state even if the CT tail is truncated, recapitulating the functional features of the intact protein \cite{lek2017swapping}. Several biochemical and biophysical studies have focused their attention on understanding the structure and conformation of the lipid-associated state of apoE3 as complexes with phospholipids such as 1,2-dimyristoyl-sn-glycero-3-phosphocholine (DMPC) or 1-palmitoyl-2-myristoyl-sn-glycero-3-phosphocholine (POPC): Fluorescence resonance energy transfer (FRET) analysis indicated that the NT domain helix bundle of lipid-free apoE3 (apoE3-NT) undergoes a lipid-triggered conformational change involving movement of the two domains away from each other \cite{narayanaswami2001lipid} and opening of the helix bundle \cite{fisher2000lipid} to reveal a hydrophobic continuum that facilitates interaction with lipids. Attenuated
Total Reflectance Fourier Transformed Infrared (ATR-FTIR) analysis of nanodiscs composed of DMPC/apoE3(1-191) or DMPC/apoE(201-299) revealed that the protein adopts an extended helical conformation that wraps around the phospholipid bilayer with the helical axes oriented perpendicular to the bilayer normal of the fatty acyl chains \cite{raussens1998low,raussens2005orientation}. The perpendicular orientation of the lipid-associated apoE3-NT in discoidal particles was confirmed by fluorescence parallax depth quenching analysis \cite{narayanaswami2004helix}. Further, studies using electron paramagnetic resonance spectroscopy focusing on residues 161 to 191, which appear to be unstructured in the X-ray structure of lipid free NT domain \cite{wilson1991three,wilson1994salt,dong1994human}, revealed that it adopted a helical structure in lipid-associated state in nanodiscs \cite{gupta2006lipid}.

Despite the intense focus, there are several unanswered questions and uncertainties regarding the structure and orientation of apoE in lipid associated state, which are attributed to the experimental limitations imposed by the large particle size of HDL, the intrinsically flexible nature of the protein and its ability to form a wide range of stable lipoprotein complexes of varying diameters.
To address these issues, the current study employed a computational approach, consisting in coarse-grained (CG) and all-atom (AA) molecular dynamics simulations (MDS), to understand the theoretical limits that dictate the formation of binary protein-lipid complexes of apoE3-NT and phospholipids.
A similar approach was used to probe the size, shape and relative orientation of apolipoprotein AI (apoAI) \cite{siuda2015molecular}, another apolipoprotein that bears no sequence similarity to apoE3 but belonging to the superfamily of exchangeable apolipoproteins.
The model that was developed for apoAI provides insight on the size, shape, and orientation of the discoidal nanodisc.
In this study, we employed MDS to study the lipid- associated state of apoE3-NT residues 1-183, starting with the high-resolution NMR-structure of this domain in lipid-free state (PDB ID: 2KC3) \cite{sivashanmugam2009unified}. Previous findings from experimental
approaches informed, defined, or provided the rationale for the set-up for the MDS. Using this approach, we report a model of nanodiscs comprising apoE3-NT and DMPC based on calculations of favorable conformations.

\section*{Methods and Computational Details}
\subsection*{General procedure}
We constructed initial models of the apoE3-NT nanodiscs of different sizes, determined by a variable number of DMPC molecules. Specifically, we focused on nanodiscs of 240--420 DMPC molecules (120-210 DMPC molecules per leaflet). In preliminary calculations, most systems with less than 240 and more than 420 DMPC molecules were not stable during MDS. The coordinates of the initially prepared structures were propagated in time using MDS, first using the CG method. Then, the last structure of each CG simulation was converted to an AA model, which was subsequently propagated in time using MDS. The trajectories of the CG and AA simulations were analyzed with respect to the following properties: a) nanodisc diameter, b) radius of gyration, c) solvent accessible surface area (SASA), d) root-mean-square deviation (RMSD) in the positions of the protein chains, e) protein contacts, and f) DMPC order parameter. Based on these quantities, we evaluate trends and gauge the overall stability of the nanodiscs.
\subsection*{Computational Details}
\begin{figure}
    \centering
    \includegraphics[scale=0.4]{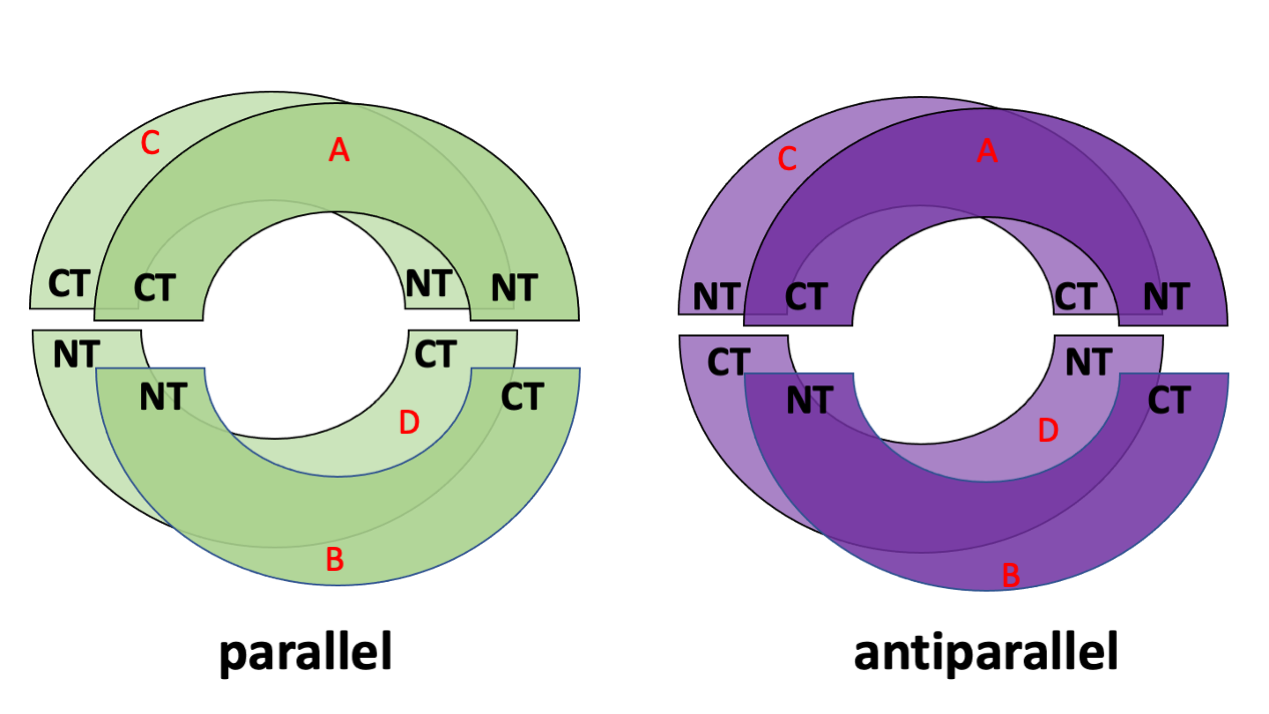}
    \caption{Schematic illustration of the initial assembly of apoE3-NT configuration in parallel (left) and antiparallel (right) protein configuration. N-terminus (NT) and C-terminus (CT) are indicated for each chain. Red letters indicate the chain label. }
    \label{fig:parallel_anti}
\end{figure}
To set-up the initial structures of apoE3-NT/DMPC nanodiscs, a protein chain was generated by unwinding the tertiary structure of the 4-helix bundle structure of the 2KC3 PDB-structure of apoE3-NT \cite{sivashanmugam2009unified}. With exception of the loops connecting the helices, the secondary structure was kept intact. Two identical protein chains were then arranged in a circular formation, 0.5~nm, apart in tandem. This system was duplicated and placed 3 nm below the initial system, yielding four total apoE3-NT protein chains.
This was done in a parallel or antiparallel fashion (Figure~\ref{fig:parallel_anti}). The energy of these systems was minimized with 10,000 steepest descent and 5,000 conjugate gradient optimization steps. After energy minimization of the protein portion, the structures were converted to a CG model using the martinize script \cite{de2013improved}. Subsequent steps follow Tieleman's protocol of apoA1, regarding assembly and MDS \cite{siuda2015molecular}.
Using the CG parallel and antiparallel structures, CG models of DMPC molecules \cite{jambeck2012derivation} were added inside the circular protein arrangement, forming a bilayer. This was accomplished using the bio.b-gen building tool \cite{biobgen}. The protein/DMPC system was embedded with CG water molecules and counterions were added to achieve charge neutrality.
Following CG-MDS, the final frame of every system was converted to an AA model for further MDS sampling.  We studied systems with 240-420 DMPC molecules per system, assembled both in parallel and antiparallel protein configurations, increasing the number of DMPC by 20 (20 total assembled nanodisc systems).
Systems are labeled by the number of DMPC molecules, hyphenated with the letter P or A, for parallel and antiparallel protein configuration, respectively (for example, 240-A denotes a nanodisc with 240 DMPC molecules and the apoE3-NT in antiparallel configuration). All MDS and energy minimizations were performed using GROMACS (2019 version) \cite{bekker1993gromacs,van2005gromacs,hess2008gromacs,pronk2013gromacs}.

\subsubsection*{Coarse Grain Molecular Dynamics Simulations} The Martini CG model (version 2.2) for proteins and lipids was used for this study \cite{marrink2004coarse,marrink2007martini,monticelli2008martini,ingolfsson2014lipid,periole2009combining}. The energy of the assembled nanodiscs was minimized using 10,000 steepest descent optimization steps. Three equilibration steps were then performed: for a) 1 ns with a time step of 10 fs, for b) 4 ns with a 20 fs time step and with lipid molecules frozen in the z-direction, and c) for 5 ns with a 20 fs time step without any constraints. All subsequent simulations were propagated for 2 $\mu s$ using a 20 fs time step, while the neighbor list was updated every 10 steps. Over the three steps, the Lennard-Jones potential was shifted to zero between {0.9} -- {1.2} nm, and the Coulomb potential was shifted to zero between 0 and 1.2 nm with a relative permittivity constant of $\epsilon_r$ = 15. An isotropic pressure coupling scheme of 1 bar was used for all simulations with a compressibility of 3$\times$10$^{-4}$ bar$^{-1}$. In step c), the pressure was relaxed using the Berendsen barostat \cite{berendsen1984molecular} with a relaxation time constant of $\tau_P$ = 5.0 ps; subsequently, a Parrinello-Rahman barostat \cite{parrinello1981polymorphic} was
used with $\tau_p$ = 12.0 ps. The target temperature was kept at 310 K using a Berendsen thermostat \cite{berendsen1984molecular} with a characteristic time $\tau_T$ = 2.0 ps.

\subsubsection*{All-Atom Molecular Dynamics Simulations}
The protein and DMPC part of the CG nanodisc systems (without counter ions and water molecules) were converted from CG to the AA Amber representation using the backward.py script \cite{wassenaar2014going}. The resulting AA models were then centered in a cubic box, and explicit ions and TIP3P \cite{jorgensen1983comparison} water molecules were added using GROMACS. AA calculations employ the The CHARMM36 force field \cite{huang2017charmm36m}. Before MDS, the energy of the structures were first minimized by 10,000 steps of steepest descent to remove any backbone strain resulting from the conversion from CG to AA. Then, the systems were equilibrated by NVT MDS for 2 ns using a 1 fs time step, followed by NPT equilibration for 1 ns using a 1 fs time step. After equilibration, MDS was carried out for 50 ns using  a 1 fs time step. Long-range electrostatic interactions beyond 1.0 nm were calculated using particle-mesh Ewald \cite{darden1993particle}, and van-der-Waals interactions were switched to zero for distances larger than 1.0 nm. Isotropic pressure was maintained at 1 bar, with a compressibility of 3$\times$10$^{-4}$ bar$^{-1}$, using the Parrinello-Rahman barostat with a relaxation time of $\tau_p$ = 4.0 ps. Temperature was kept at 310 K using a Berendsen thermostat with a relaxation time of  $\tau_T$ =0.1 ps.

As a reference for the nanodisc systems, a pure DMPC bilayer containing 150 DMPC molecules (75 per layer) was generated using CHARMM membrane builder \cite{jo2008charmm,lee2018charmm}. The minimization and equilibration parameters used were identical to the parameters for nanodiscs stated above. The MDS was performed for 50 ns under periodic boundary conditions.

\subsubsection*{Analysis}
The final 100 ns of the CG-MDS and final 10 ns of AA-MDS were analyzed. All values reported, represent averages over 200 individual snapshot structures of these trajectories, extracted in equal time intervals (for CG every 0.005 ns, for AA every 0.05 ns).
Analysis was performed using the tools provided by GROMACS \cite{castillo2013free}, unless otherwise mentioned.
 Snapshot structures were reoriented such that the bilayer normal was parallel to the z-axis.

The phosphate ion mass density distributions were measured along the z-axis divided into 100 bins and symmetrized in the same manner for the lipids and proteins. Nanodisc bilayer thickness was determined from the peak to peak distance of the phosphate ion mass density distributions.

We defined the nanodisc diameters by assuming an elliptical nanodisc shape and measuring the diameters of both the major (x-axis) and minor (y-axis) axis, using the atomic mass density distributions.
Density distributions were symmetrized around the center of mass of the nanodisc for consistent measurements of the diameters in the x and y directions.
To obtain a more detailed description, we determined the diameters
 by measuring the distance between the left and right outer maxima of the protein mass density distribution, and in addition, by measuring the distance between the positions to the left and right of the center, at which the DMPC density distribution decayed to 1 kg/m$^3$.
The density distribution of the snapshot structures was determined by binning both, x- and y-directions into 100 intervals.

The average diameter was determined as mean value of the diameter along the x-axis and along the y-axis, determined from the mass density distribution.
The SASA of the CG nanodiscs was calculated using a radius of 0.265 nm for the water probe particles. For the AA system, the SASA was calculated using a radius of 0.14 nm for the water probe particles. Both calculations employ 48 dots per sphere.

To describe the structural organization of the protein portion, contacts of amino acids between different protein chains within 1 nm were extracted from 200 structures of the final 10 ns of the AA-MDS.
In our further analysis we only consider protein-protein contacts below 0.4 nm.

Order parameters indicate the degree of ordering in the hydrocarbon tails of phospholipids in a membrane \cite{Douliez_1998,Vermeer_2007,Smondyrev_1999a}.
The S\textsuperscript{CD} bond order parameter describes the orientation of a C-H bond relative to the bilayer normal and can easily be extracted from simulations \cite{Tu_1998,Smondyrev_1999b,Hofsass_2003}.
For the $i$th carbon along the lipid tail, it is defined as
\begin{equation}\label{s_cd}
S_i^{CD} = \frac{1}{2}\langle 3\cos^{2}(\theta_i) - 1 \rangle  ,
\end{equation}
where $\theta_i$ is the angle between the C-H bond of the $i$th carbon atom and the bilayer normal, brackets indicate time and ensemble average.
Order parameters were averaged over Sn1 and Sn2 chains of the DMPC molecules.
Order parameters were determined for every AA nanodisc system and for the pure DMPC bilayer using the VMD extension MEMBPLUGIN \cite{guixa2014membplugin}. For every nanodisc system, we calculated order parameter in three different ways: a) Values were calculated using all DMPC molecule in nanodisc (full), b) using a core sample from every nanodisc system that contained all DMPC molecules within three nm of the lipid center of mass (core), and c) for a rim section, containing all DMPC molecules with 14 nm from the protein belt (rim).

\section*{Results and Discussion}
\subsection*{Coarse Grain Simulations}
\begin{figure}
\includegraphics[scale=0.3]{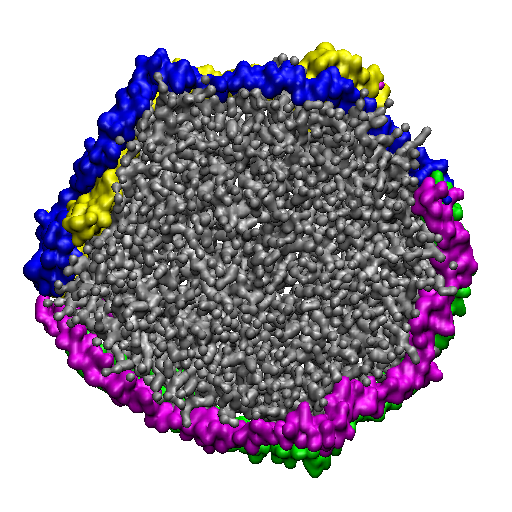}
\includegraphics[scale=0.3]{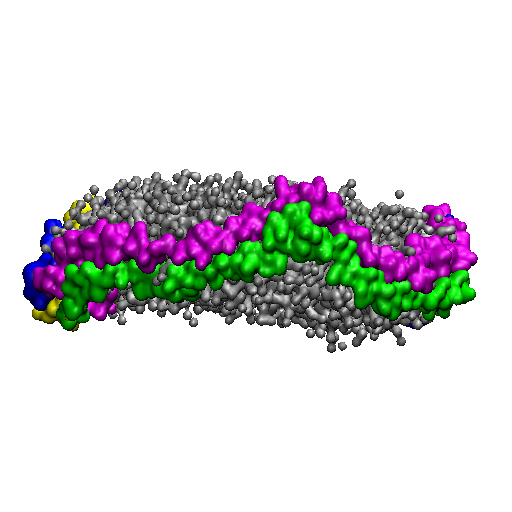}
\includegraphics[scale=0.3]{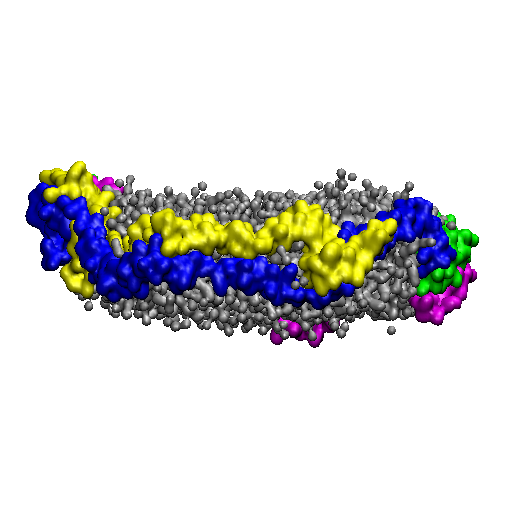}
\includegraphics[scale=0.3]{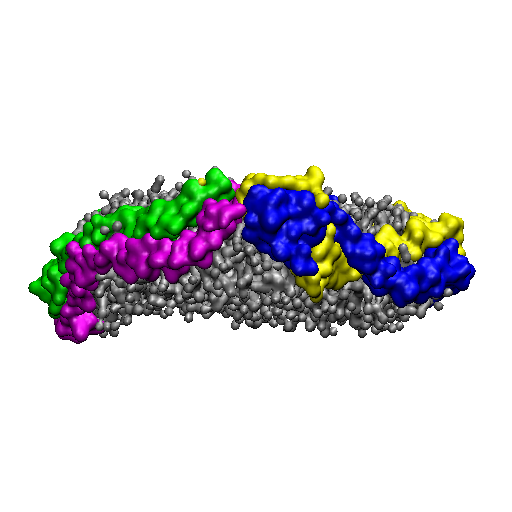}
\includegraphics[scale=0.3]{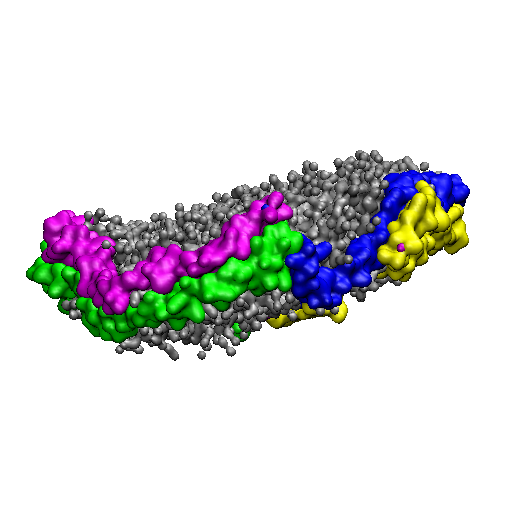}
\caption{\label{cg_anti180}Five different views of a representative nanodisc, obtained as the last structure of the CG-MDS of the 360-A system. DMPC molecules are represented in grey. Protein chains A, B, C, and D are represented in blue, magenta, yellow, and green ribbons, respectively. }
\end{figure}

Analyzing the trajectories of the CG-MDS, we note that the system with 240 DMPC molecules and parallel protein configuration (240-P) did not lead to a stable nanodiscs. The remaining systems all lead to stable nanodiscs with typical belt-like disc structure, as shown in Fig.~\ref{cg_anti180} for 360-A, as a representative system. In a few simulations (400-P, 240-A, 260-A, 400-A) individual lipid molecules escaped from the nanodisc cluster. Visualizations of the last frame of all CG-MDS are given in the Supporting Information (SI).

\begin{figure}
    \centering
    \includegraphics[scale=0.4]{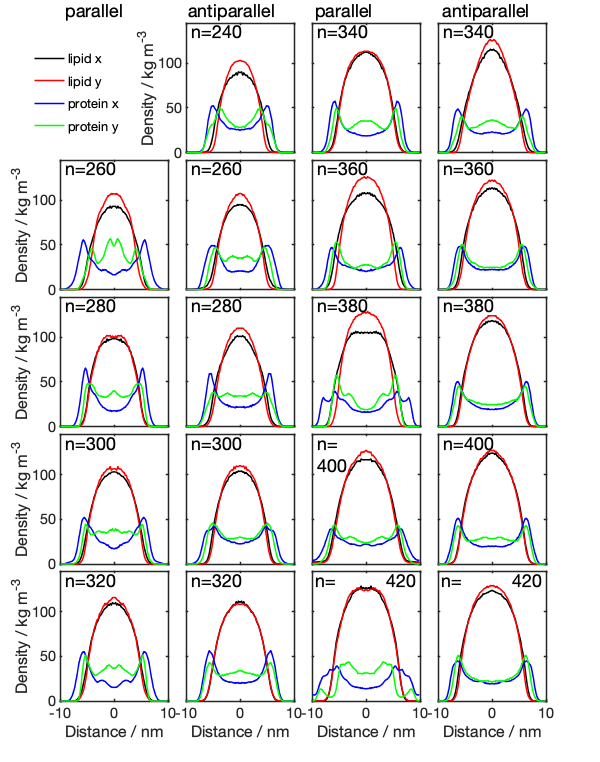}
    \caption{Density distributions across the membrane for parallel (1st and 3rd column) and antiparallel (2nd and 4th column) nanodiscs with different number of DMPC molecules ($n$), obtained from CG simulations. The center of the membrane is located at the origin of the coordinate system. Lipid x and y, denoted the density distribution obtained from the lipid molecules in x- and y-direction, respectively; protein x and y, denotes the density distribution obtained from the protein density distribution, in x- and y-direction, respectively.}
    \label{fig:densities_cg}
\end{figure}

\begin{figure}
    \centering
    \includegraphics[scale=0.4]{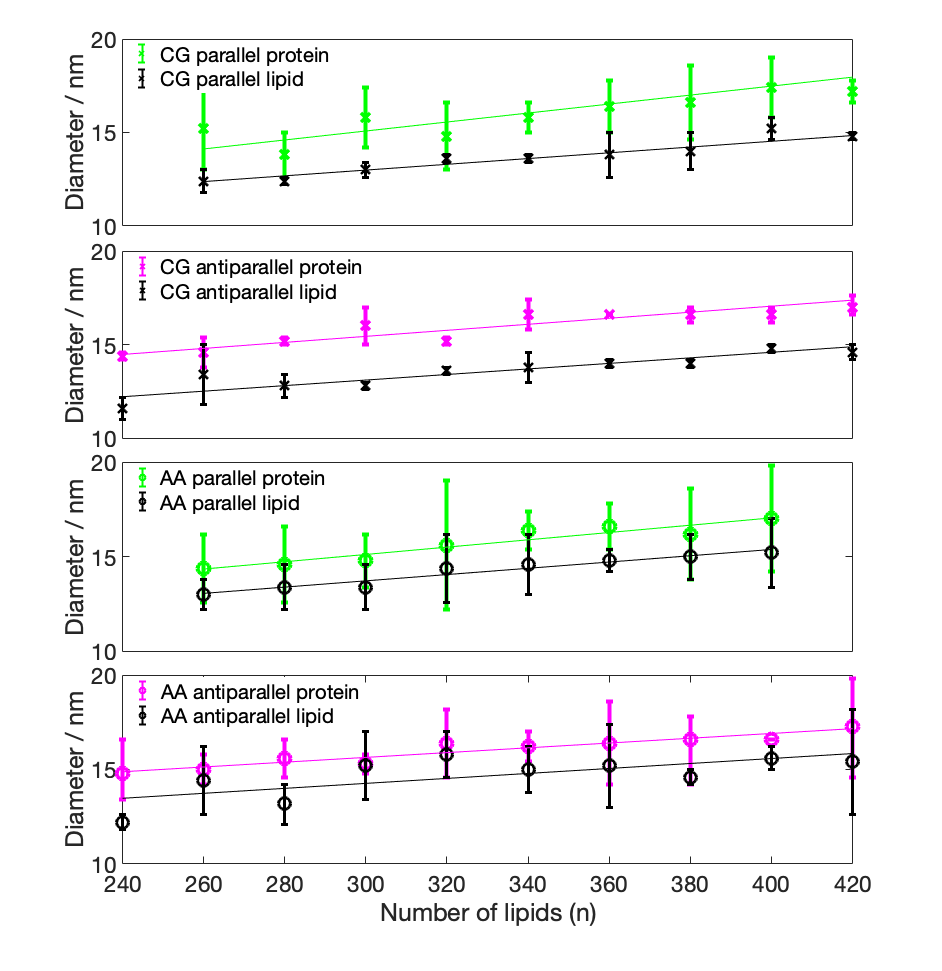}
    \caption{Average diameter for the CG and AA nanodiscs, measured based on the protein density (color) and based on the lipid density (black) distribution. Lower limit of the errorbars indicate minor axis, upper limit indicates major axis, symbols indicate the average between both axis. Linear regression is also given.}
    \label{fig:diameter}
\end{figure}

The mass density distributions for the nanodiscs obtained from CG simulations (Fig.~\ref{fig:densities_cg}) showed that some of the nanodiscs exhibit almost perfect circular shape, which is indicated by a very similar lipid density distribution in x- and y-direction. Examples of this behavior are systems 340-P, 400-A, 320-A, and 420-P. The remaining discs exhibit a more elliptical shape, with different diameter in x- and y-direction, indicated by the errorbars in Fig.~\ref{fig:diameter} (values given in Table S1, SI). Diameters obtained from the protein density range from 13.8 to 19.8 nm. Examining the diameter as a function of number of lipid molecules, we find a linear relationship. We further notice that the diameter predicted by using the protein density is always larger than the diameter obtained from the lipid density distribution, which confirms the double belt-like structure. Furthermore, we notice that variations in diameter along x- and y-directions are larger if they are obtained on the basis of the protein density than if they are obtained from the lipid density.

\begin{figure}
    \centering
    \includegraphics[scale=0.4]{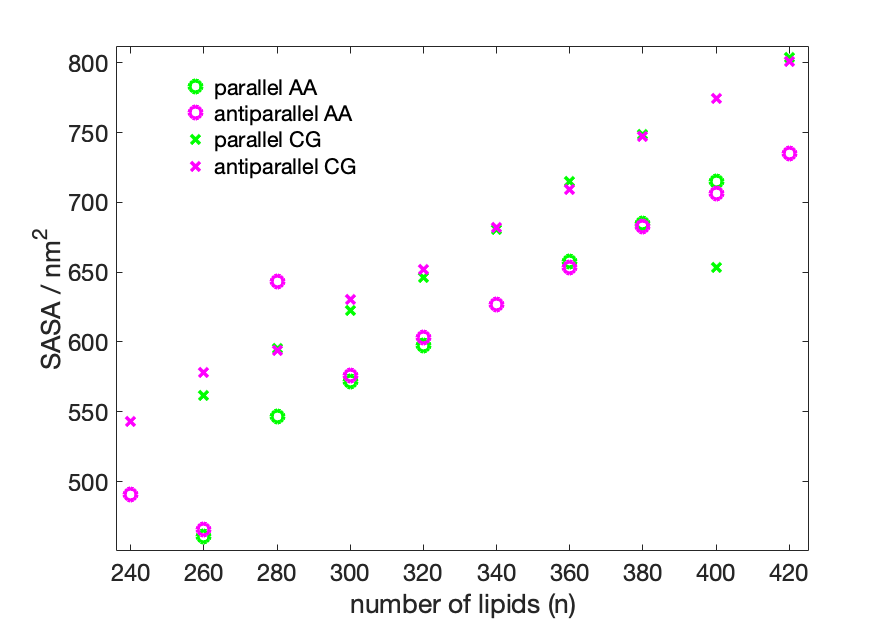}
    \caption{SASA in nm$^2$ as a function number of DMPC molecules.
     420-P could not be converted from CG to AA due to overlapping atoms.}
    \label{fig:sasa}
\end{figure}

\begin{figure}
    \centering
    \includegraphics[scale=0.4]{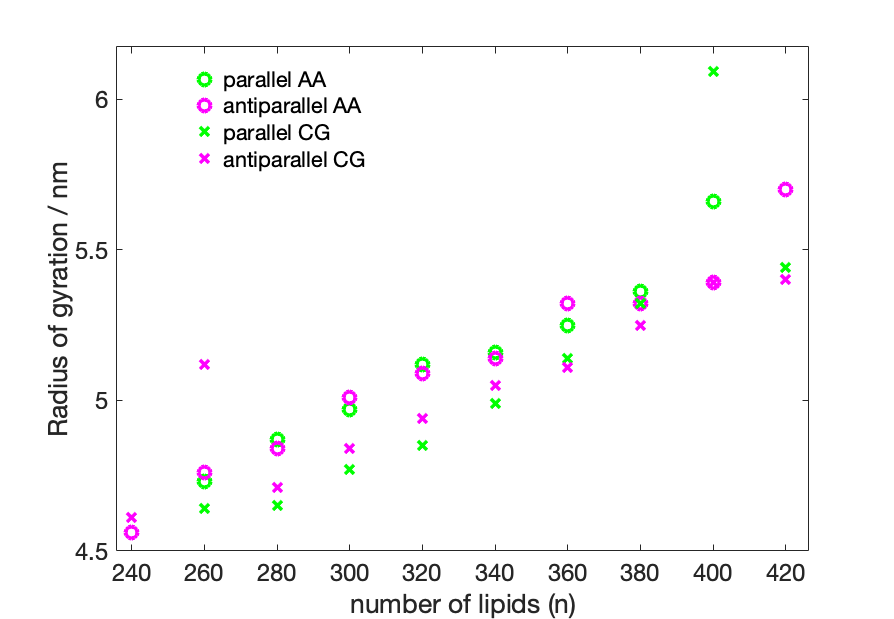}
    \caption{Radius of gyration as a function number of DMPC molecules.
     The 420-P could not be converted from CG to AA due to overlapping atoms.}
    \label{fig:radius}
\end{figure}

With the exception of the 400-P system, the SASA follows a linear increase as the number of lipid molecules is increased (Fig.~\ref{fig:sasa}, Table S2, SI).
The trend of the radius of gyration as a function of the number of lipids (Fig.~\ref{fig:radius}) also appears to be linear, with, however, less perfect behavior than the SASA and two outliers (400-P, 260-A). The two outliers are both two cases, where lipid molecules have left the nanodisc.


\subsection*{All-atom simulations}
\subsubsection*{Densities}
All systems, except 420-P, were successfully converted from CG to AA models. The 420-P structure could not be converted because there was a large atomic overlap after conversion, leading to an unstable AA system.
The phosphate to phosphate bilayer thickness of all discs amounted to 3.4 nm, which is equal to the thickness of the pure DMPC bilayer and in agreement to previous simulations \cite{barrera2017modeling}.

\begin{figure}
\includegraphics[scale=0.3]{./new1.png}
\includegraphics[scale=0.3]{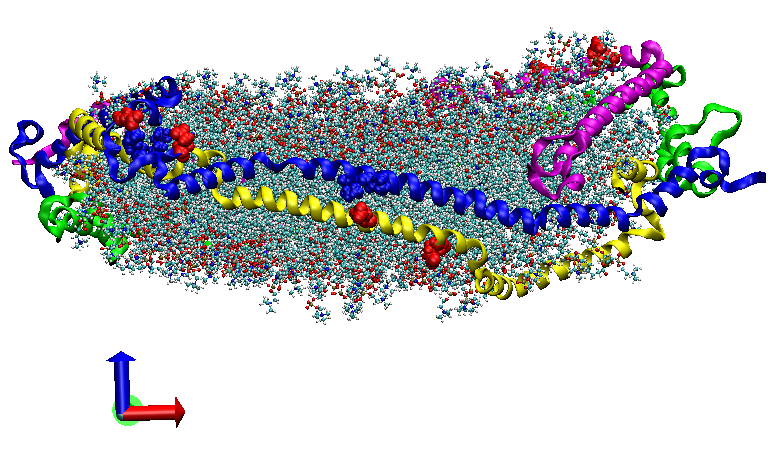}
\includegraphics[scale=0.3]{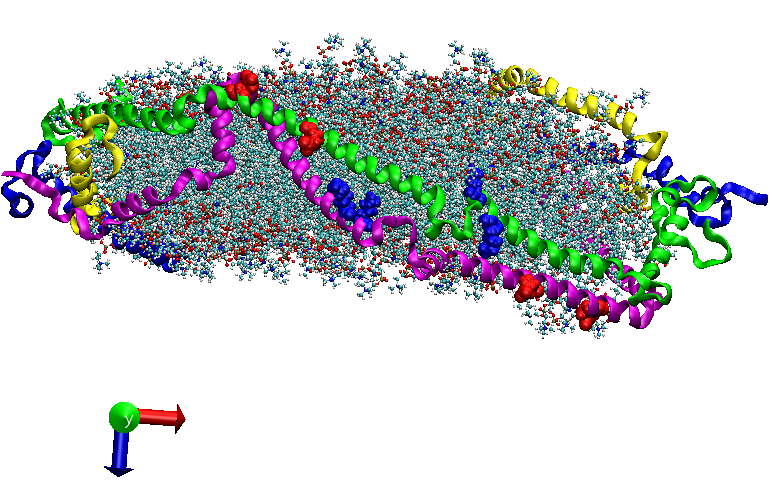}
\includegraphics[scale=0.3]{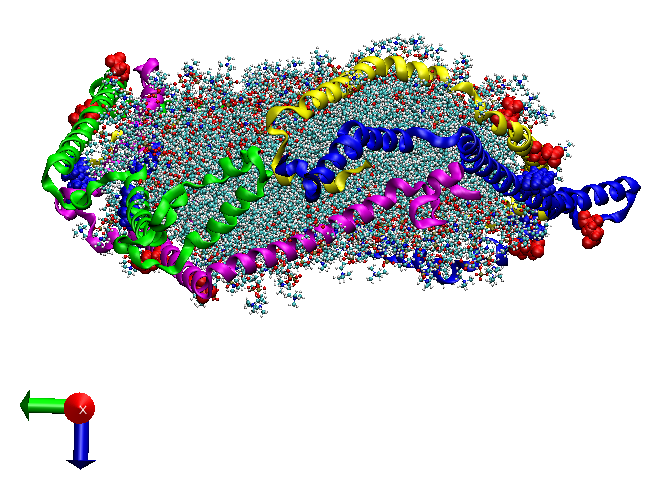}
\includegraphics[scale=0.3]{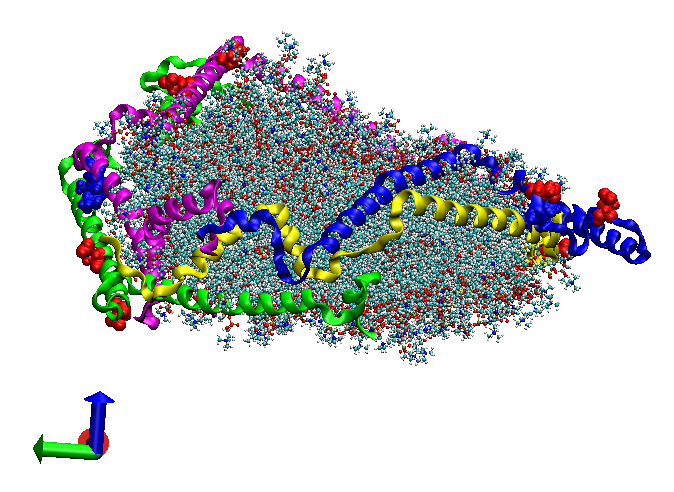}
\caption{\label{aa_structure}Five different views of a representative nanodisc, obtained as the last structure of the AA-MDS of the 280-A system. DMPC molecules are represented by the ball and stick model. Protein chains A, B, C, and D are represented in blue, magenta, yellow, and green ribbons, respectively. Amino acids K75 and K72 are represented in blue balls model; amino acids E109 and E121 are represented in the red balls model.}
\end{figure}

\begin{figure}
    \centering
    \includegraphics[scale=0.4]{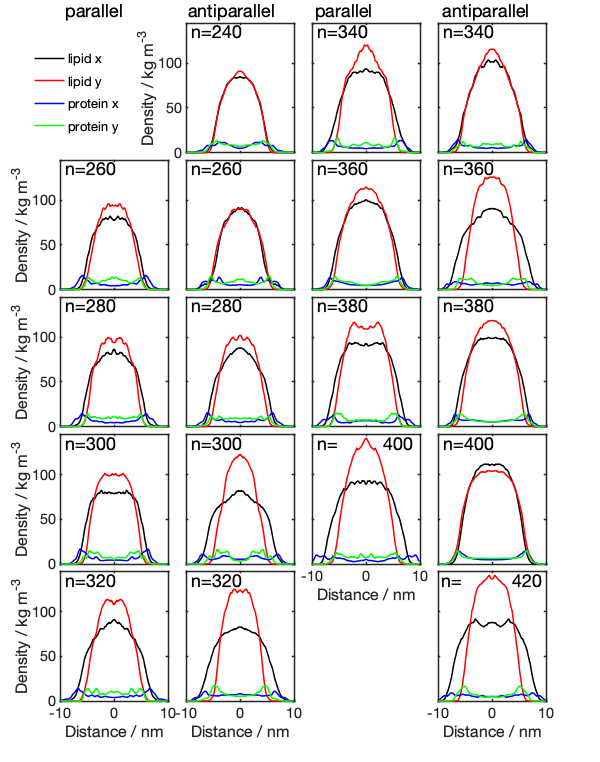}
    \caption{Density distributions across the membrane for parallel (1st and 3rd column) and antiparallel (2nd and 4th column) nanodiscs with different number of DMPC molecules ($n$), obtained from AA-MDS. The center of the membrane is located at the origin of the x-axis.
     The 420-P system could not be converted from CG to AA due to overlapping atoms. Lipid x and y, denoted the density distribution obtained from the lipid molecules in x- and y-direction, respectively; protein x and y, denotes the density distribution obtained from the protein density distribution, in x- and y-direction, respectively.}
    \label{fig:densities}
\end{figure}

Analyzing the density distribution of the AA models (Fig.~\ref{fig:densities}), we notice that compared to the CG-MDS, nanodiscs show a higher degree of asymmetry between the density distributions in x- and y-direction, deviating stronger from the circular shape.
Average nanodisc diameters obtained from AA protein mass density distribution are similar to the ones obtained from the CG models, ranging from 14.4 to 17.3 nm (Fig.~\ref{fig:diameter}, Table S3) and showing a linear trend as a function of the number of DMPC molecules.
The slope of this linear relationship is similar as in the case of the CG models.
Also, the SASA of the AA systems follows a linear increase with the number of DMPC molecules, similar, albeit slightly smaller than in CG-MDS (Fig.~\ref{fig:sasa}). The differences are possibly due to the different size of the water probe molecules in CG and AA, that were used to determine the SASA. 260-P, 260-A, and 280-A are outliers to the linear behavior.
The radius of gyration determined for the AA models appears to be slightly larger than in CG-MDS (Fig.~\ref{fig:radius})
and also shows a linear increase with increasing number of lipid molecules.

The RMSD of the protein backbone (Table~\ref{tab:table1}) range from 0.43--0.77 nm. Except for the discs with 260 and 300 DMPC molecules, the RMSD of the protein backbone is consistently smaller in the antiparallel configuration, indicating more rigid protein chains in the antiparallel configuration.
Compared to the helix bundle structure with a RMSD of 0.042 nm \cite{sivashanmugam2009unified},
RMSD values in the protein backbone of the nanodiscs are one order of magnitude larger.
\begin{table}[]
    \centering
    \begin{tabular}{c  | c c }\hline
        Number of  &  \multicolumn{2}{c}{RMSD }    \\
        DMPC &   \multicolumn{2}{c}{ protein}    \\
        molecules &  \multicolumn{2}{c}{backbone (nm)}    \\\hline
        &parallel &antiparallel \\\hline
240 & unstable    & 0.48 $\pm$ 0.10 \\
260 &  0.43 $\pm$ 0.13 & 0.58 $\pm$ 0.12\\
280 &  0.61 $\pm$ 0.12 & 0.55 $\pm$ 0.16\\
300 &  0.49 $\pm$ 0.13 & 0.55 $\pm$ 0.12\\
320 &  0.66 $\pm$ 0.21 & 0.46 $\pm$ 0.10\\
340 &  0.52 $\pm$ 0.13 & 0.43 $\pm$ 0.09\\
360 &  0.77 $\pm$ 0.24 & 0.57 $\pm$ 0.17\\
380 &  0.58 $\pm$ 0.17 & 0.57 $\pm$ 0.13\\
400 &  0.65 $\pm$ 0.13 & 0.46 $\pm$ 0.11\\
420 &    -         & 0.71 $\pm$ 0.25\\\hline
    \end{tabular}
    \caption{RMSD of the position of the protein backbone atoms of AA-MDS for all systems studied.
The 420-P could not be converted from CG to AA due to overlapping atoms.}
    \label{tab:table1}
\end{table}

\subsubsection*{Protein Contact Maps}
Protein contact maps for all AA systems are given in the SI (Figures S39-S47).
The predominant protein-protein interactions, in both parallel and antiparallel configurations, are between chain A and chain C (Blue and yellow in Fig. \ref{aa_structure}, respectively), and between chain B and chain D (magenta and green in Fig \ref{aa_structure}, respectively).  Overall, most systems in antiparallel configuration exhibit a slightly better alignment of the interacting protein chains, i.e. the sequences of interacting amino acids show less interruptions (Figures S39-S47).
In the following analysis, we only consider contacts closer than 0.4 nm.
Inspecting Table \ref{tab:contacts}, we note that antiparallel nanodiscs consistently exhibit a higher total number of contacts (140-205 vs. 51-143).
The median number of contacts is 180 $\pm$ 12 for the antiparallel configuration, which is larger than the median number of contacts for the parallel configuration (100 $\pm$ 27).
With exception of 320-A, the number of nonspecific contacts is higher than the number of polar or nonpolar contacts regardless of protein configuration.

Also the number of polar contacts is higher in the antiparallel nanodiscs.
In particular, the number of polar ionic contacts involving charged residues, R, K, H with E or D, is higher for systems with antiparallel configuration than with parallel configuration.
The number of nonpolar contacts is always larger for systems with antiparallel configuration than with parallel configurations (Table~\ref{tab:contacts}).
On average, the antiparallel nanodiscs exhibit 2.3 more nonpolar contacts than the parallel systems.

\begin{table}[]
    \centering
    \begin{tabular}{c c c c c |c  c c c c } \hline
    & \multicolumn{4}{c}{Parallel} & \multicolumn{4}{c}{Antiparallel}\\\hline
    &     polar     & nonspecific   & nonpolar   & total &  polar   &  nonspecific  & nonpolar   &  total    \\
    & total / ionic &    &    &      & total / ionic    &    &    &      \\\hline
240 &  -  / -       & -  & -  & -    &  51 / 13 & 93 & 61 & 205  \\
260 &  30 / 5       & 38 & 33 & 101  &  32 / 18 & 61 & 47 & 140  \\
280 &  8  / 1       & 27 & 16 & 51   &  41 / 14 & 73 & 56 & 170  \\
300 &  19 / 1       & 37 & 19 & 75   &  61 / 16 & 65 & 63 & 189  \\
320 &  27 / 6       & 47 & 16 & 90   &  44 / 10 & 70 & 71 & 185  \\
340 &  38 / 1       & 66 & 39 & 143  &  40 / 9 & 85 & 66 & 191  \\
360 &  32 / 9       & 59 & 33 & 124  &  57 / 11 & 78 & 57 & 192  \\
380 &  27 / 5       & 64 & 27 & 118  &  58 / 18 & 89 & 55 & 202  \\
400 &  28 / 5       & 57 & 16 & 101  &  53 / 10 & 67 & 50 & 170  \\
420 &  -  / -       & -  & -  & -    &  43 / 8  & 63 & 59 & 165  \\\hline
    \end{tabular}
    \caption{All stacking contacts between chains A and C,
between chains B and D. Contacts are defined when two side chains are closer than 4 \AA\, in distance. Stacking contacts are categorized as polar, nonpolar, or nonspecific interactions. For polar contacts, ionic contacts of R, K, or H residues paired with charged D or E residues are given in addition. The 420 parallel DMPC could not be converted from CG to AA due to overlapping atoms.}
    \label{tab:contacts}
\end{table}

Since ionic contacts between oppositely charged amino acids are expected to lead to larger interaction energies than remaining non-bonded interactions,
we interpret the larger number of ionic interactions as an indicator of higher stability of the antiparallel configurations compared to the parallel configuration. Previous studies on protein/lipid nanoparticles \cite{mei2011crystal,xu2023reconfigurable} have shown that these interactions are crucial for nanodisc stability.

\begin{figure}
    \centering
    360 DMPC molecules\\
    \includegraphics[scale=0.18]{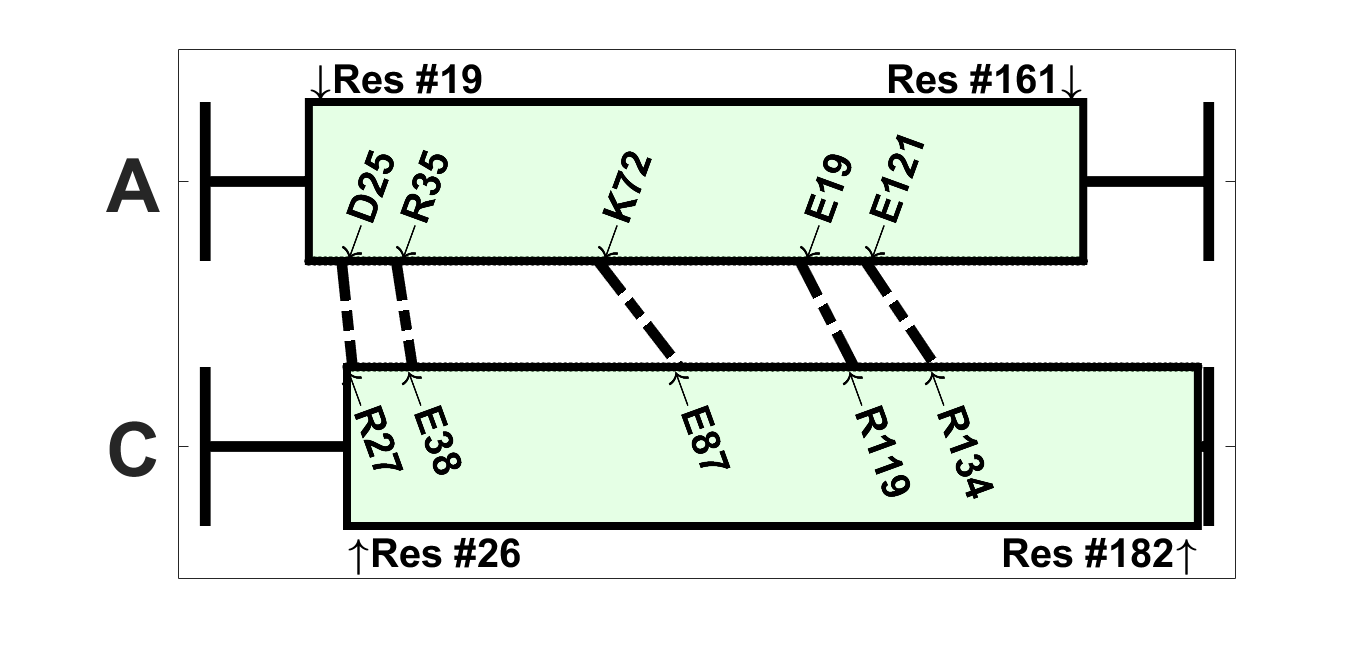}
    \includegraphics[scale=0.18]{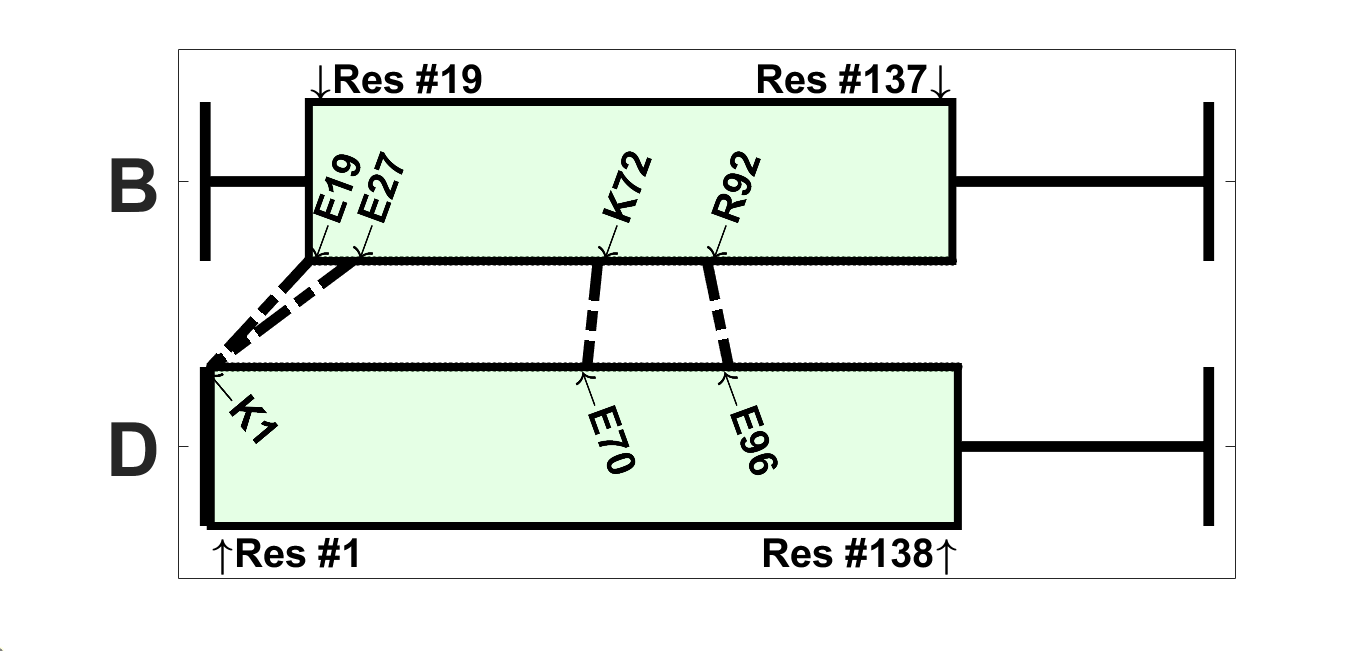}

    380 DMPC molecules\\
    \includegraphics[scale=0.18]{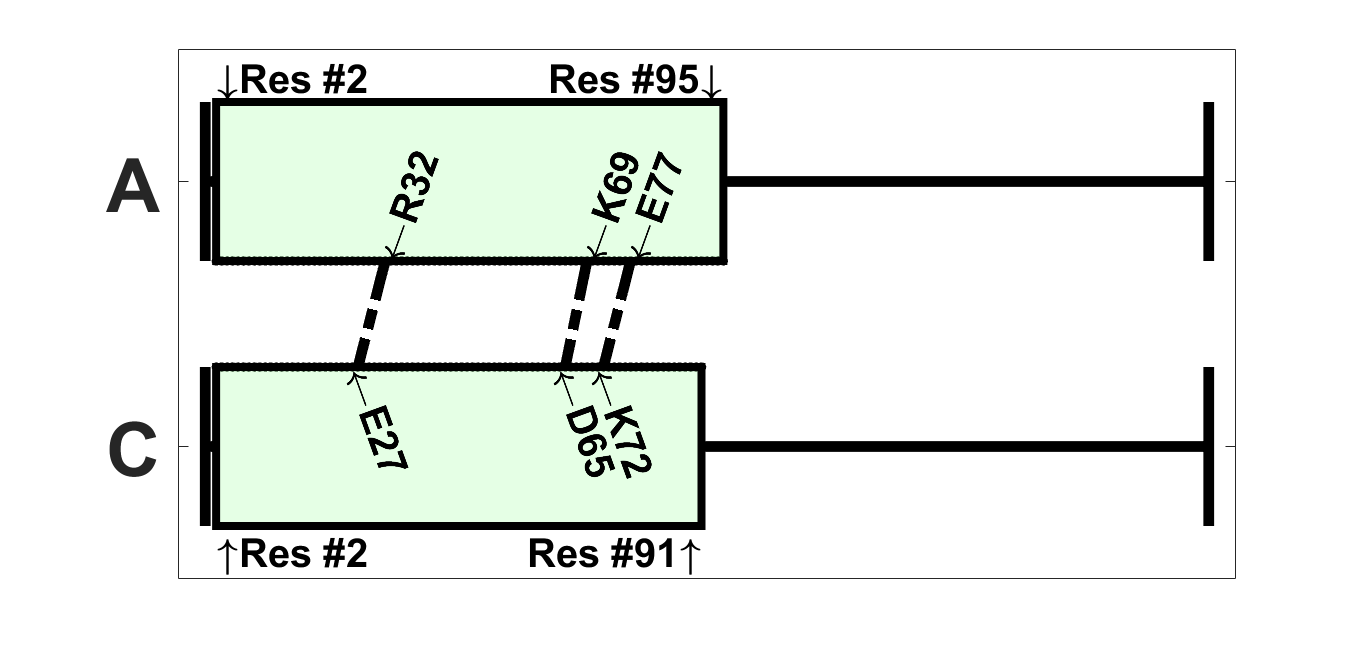}
    \includegraphics[scale=0.18]{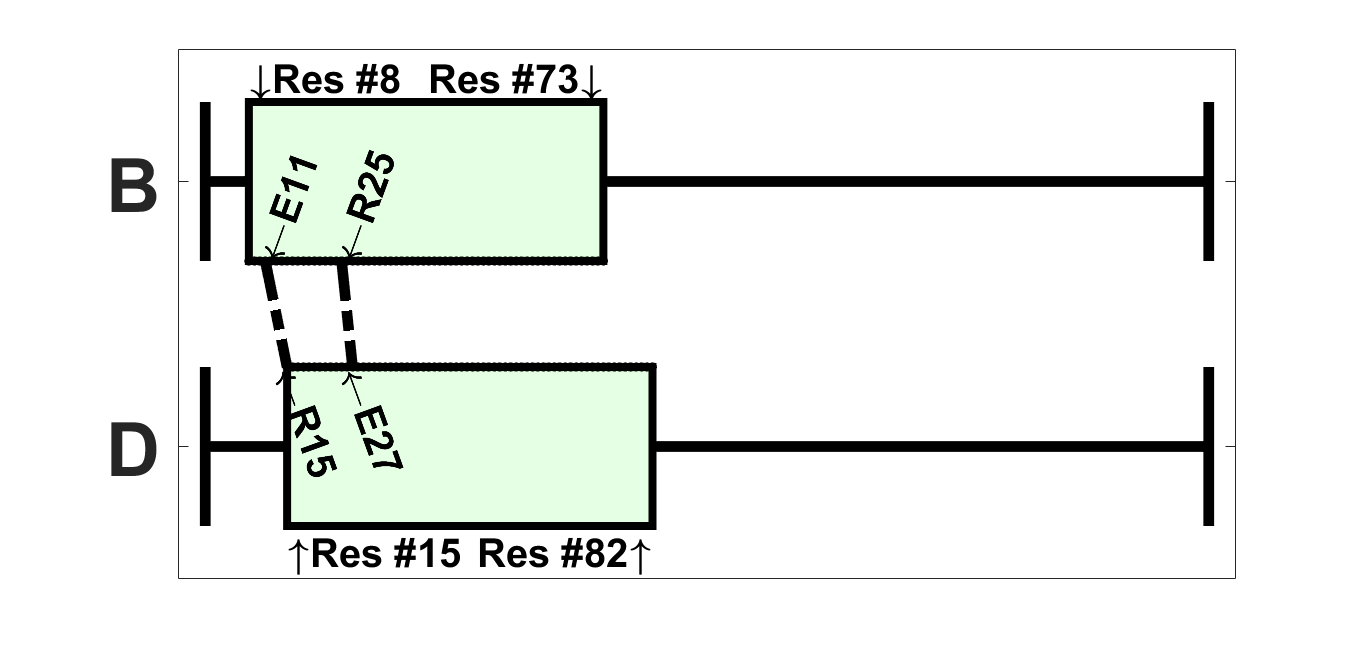}

    400 DMPC molecules\\
    \includegraphics[scale=0.18]{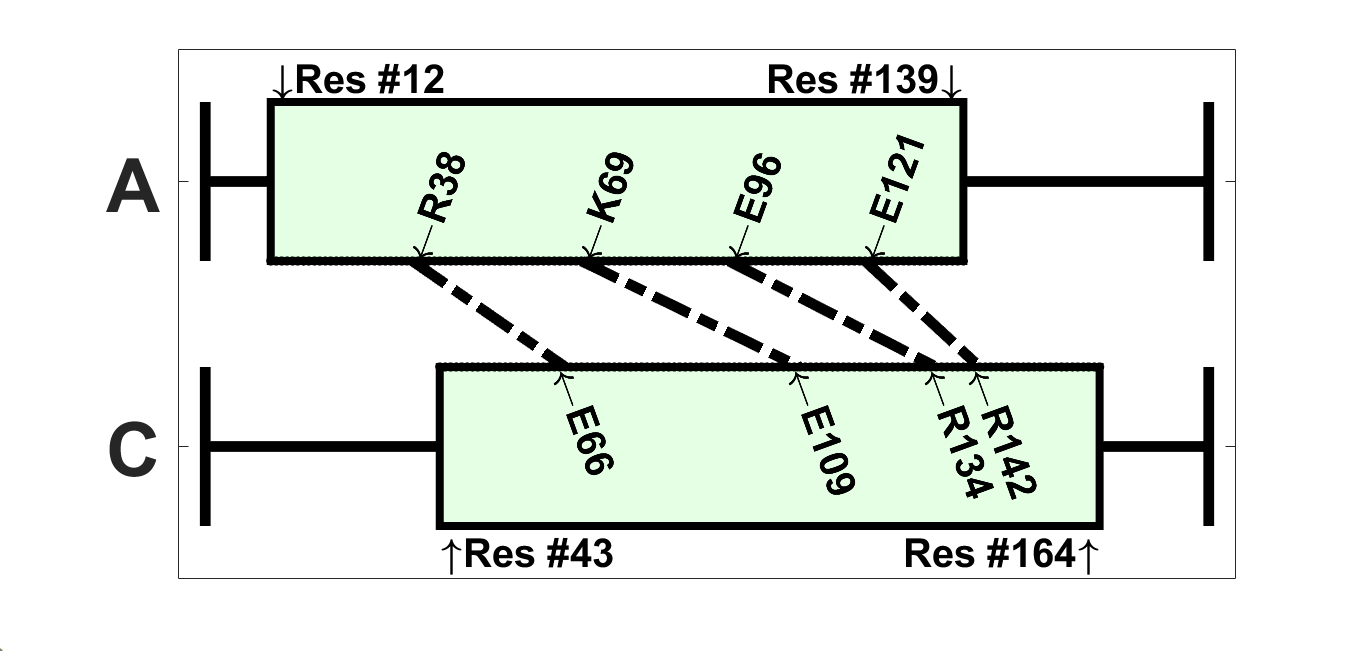}
    \includegraphics[scale=0.18]{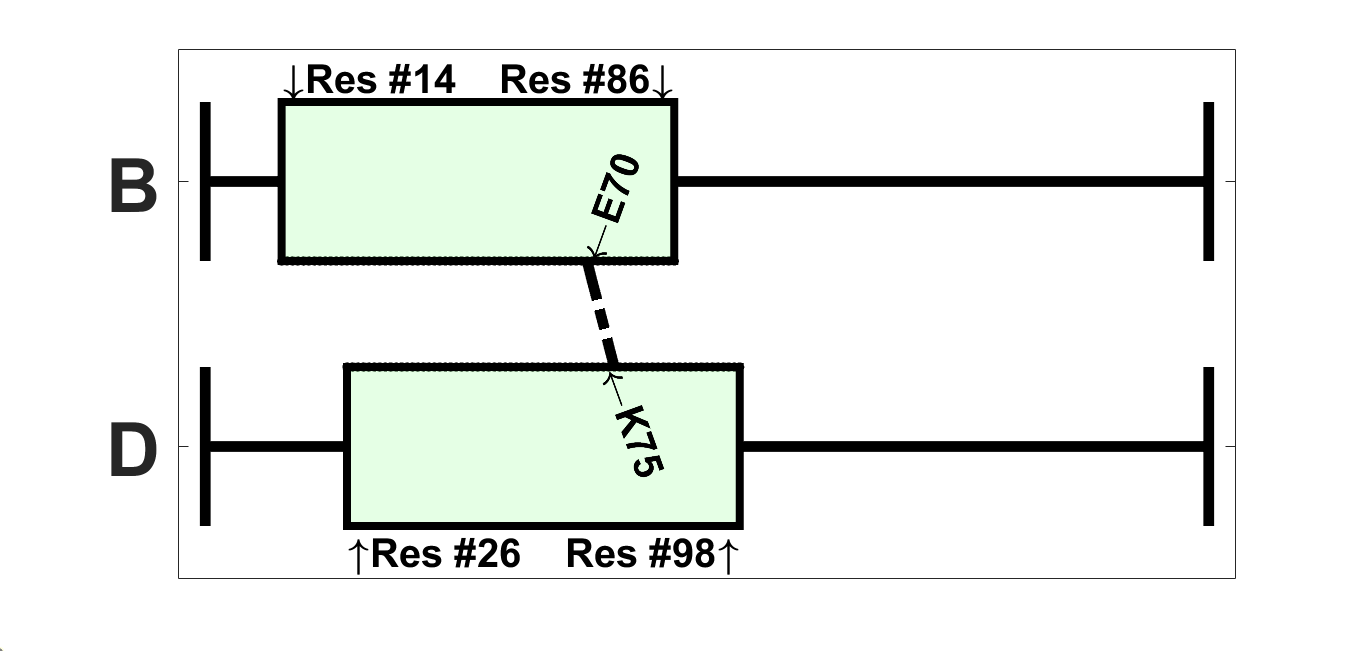}

    \caption{Visualization of the ionic contacts in nanodiscs with parallel belt configuration, shown for systems between 260 -- 400 DMPC molecules. The number of contacts is given in Table \ref{tab:contacts}.}
    \label{fig:contacts_1}
\end{figure}

\begin{figure}
    \centering
    340 DMPC molecules\\
    \includegraphics[scale=0.18]{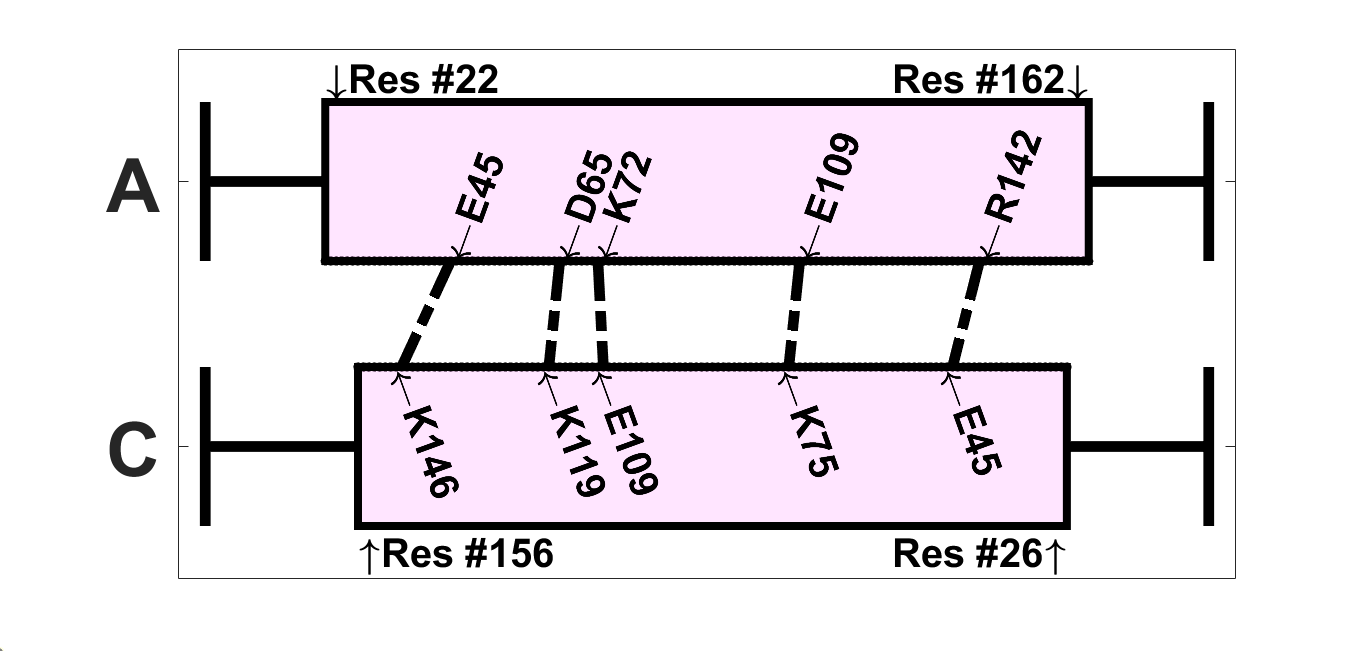}
    \includegraphics[scale=0.18]{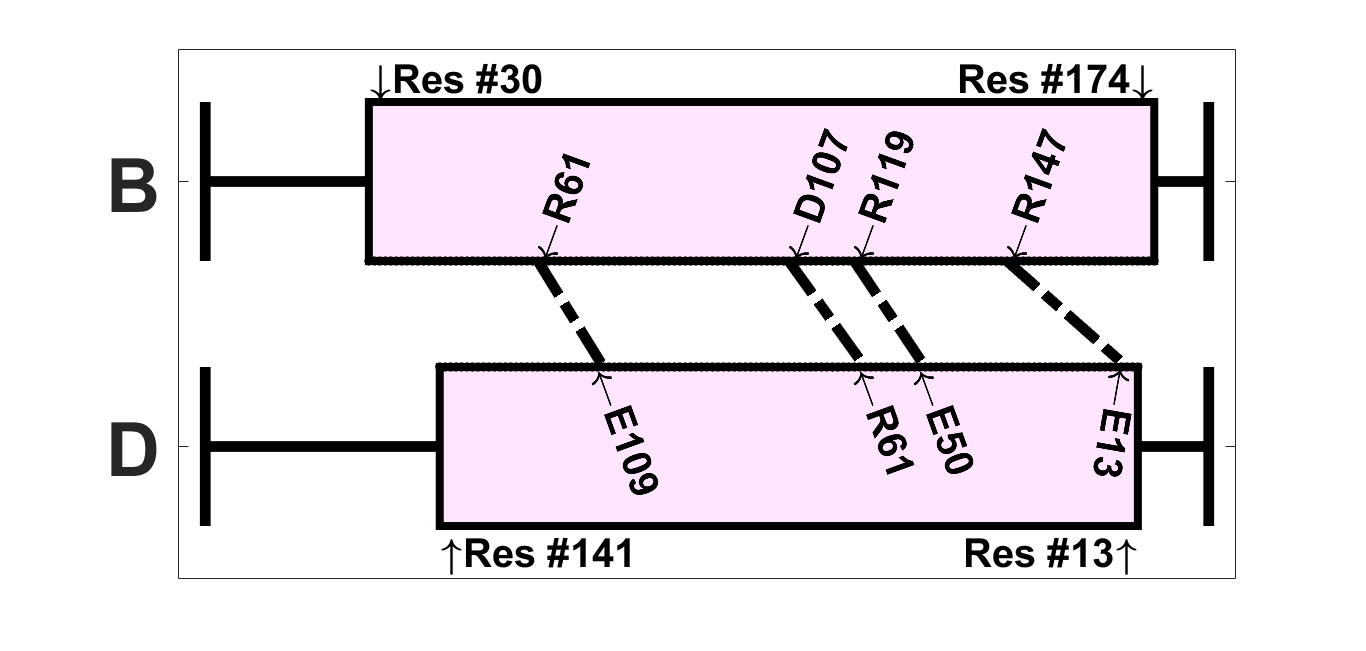}

    360 DMPC molecules\\
    \includegraphics[scale=0.18]{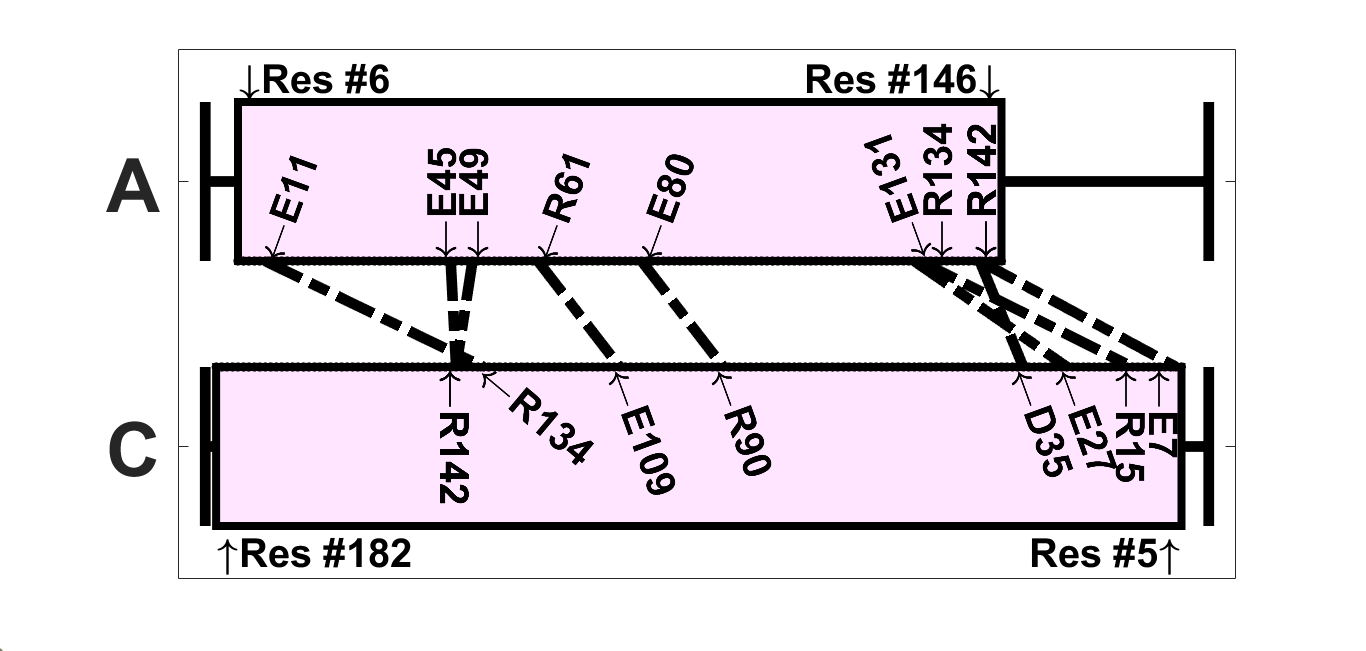}
    \includegraphics[scale=0.18]{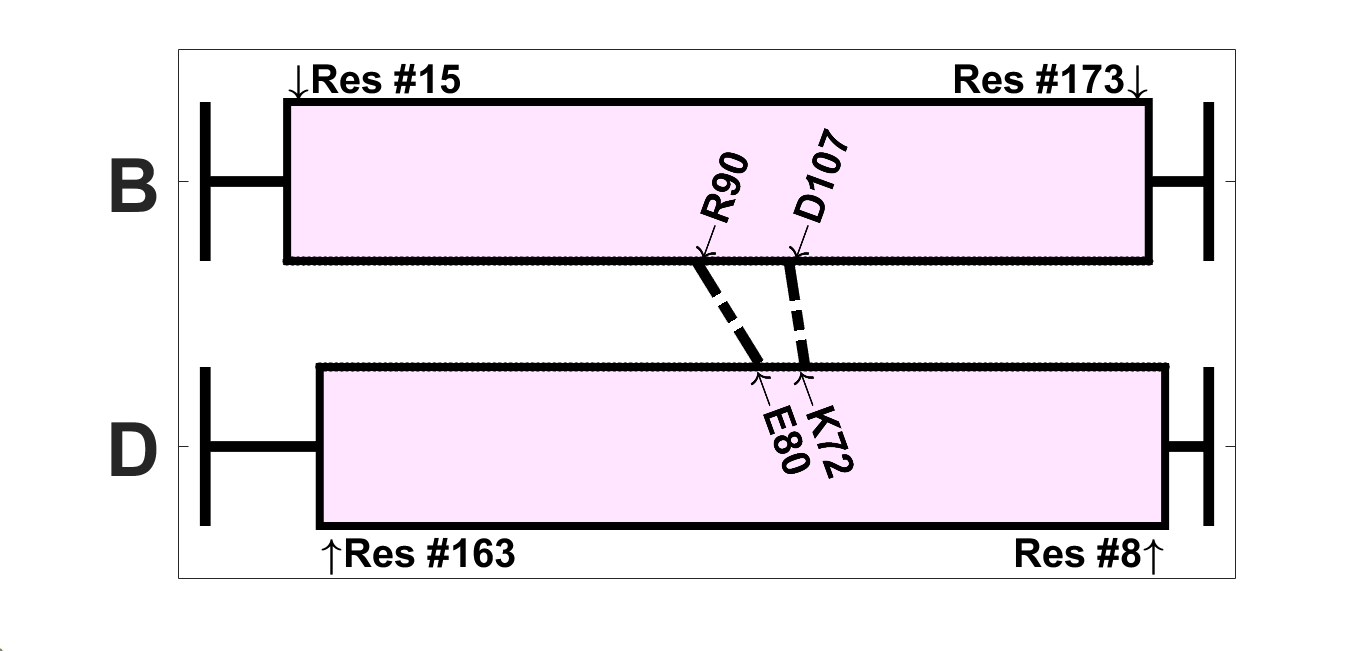}

    380 DMPC molecules\\
    \includegraphics[scale=0.18]{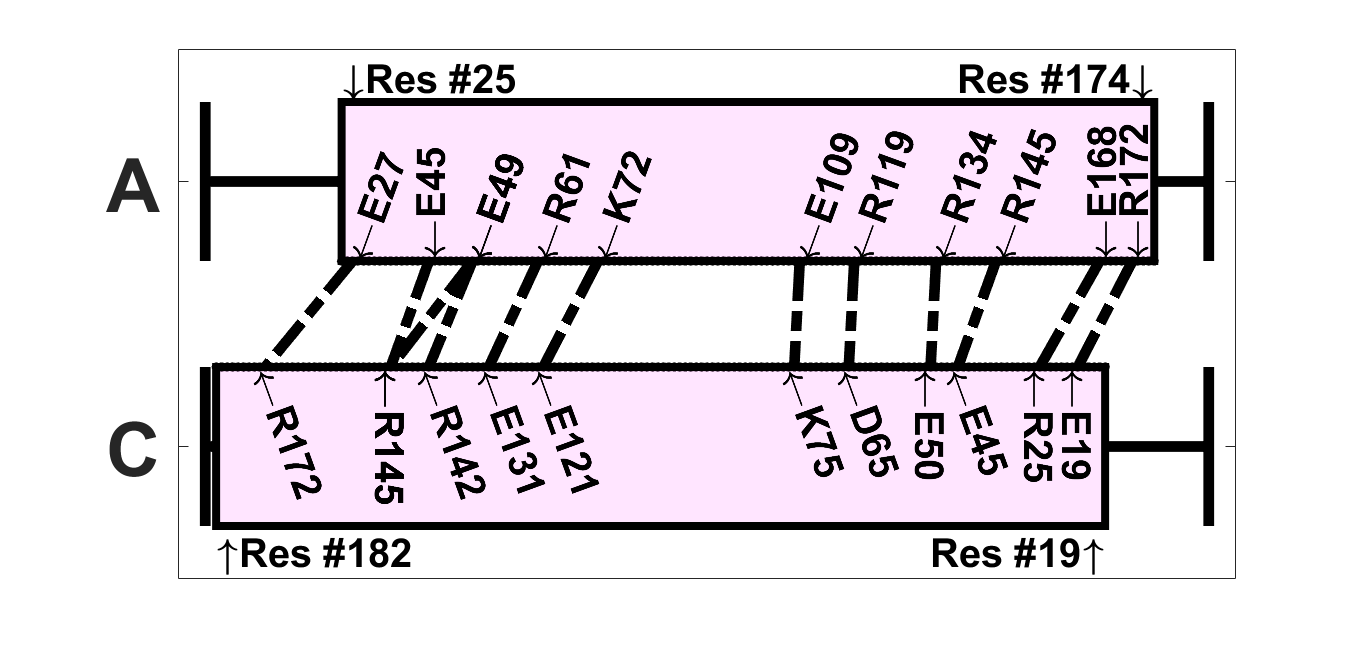}
    \includegraphics[scale=0.18]{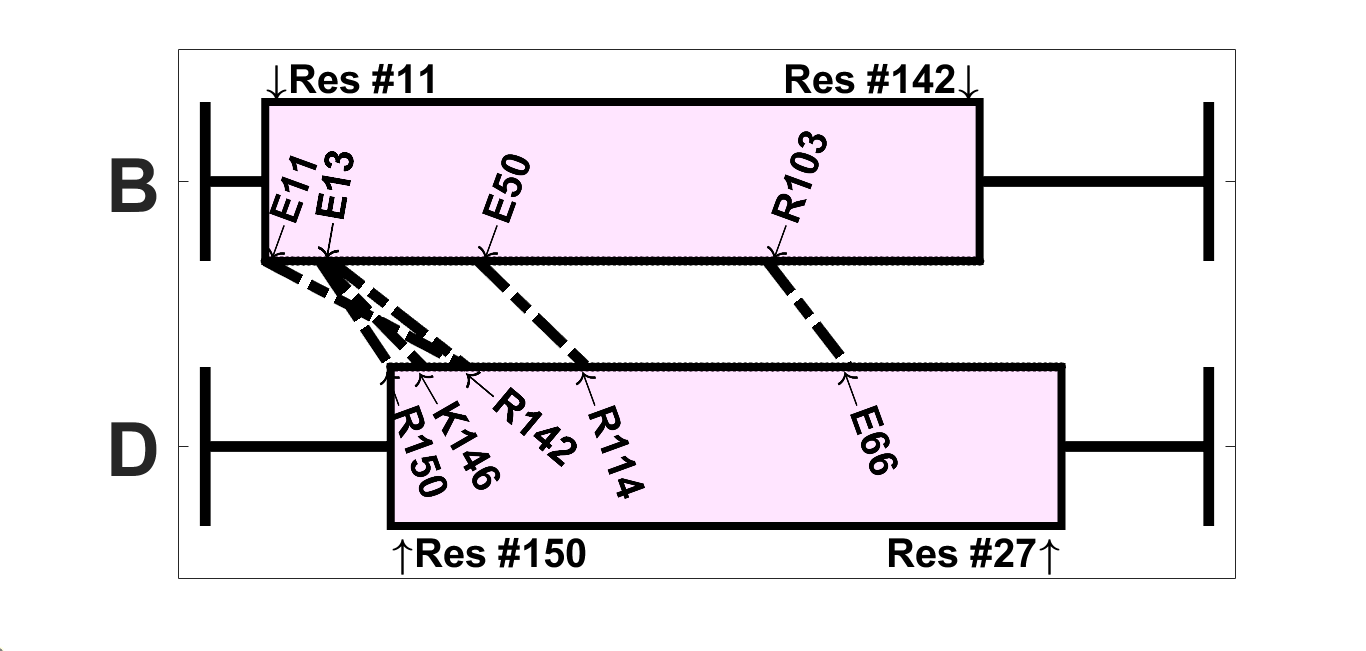}

    400 DMPC molecules\\
    \includegraphics[scale=0.18]{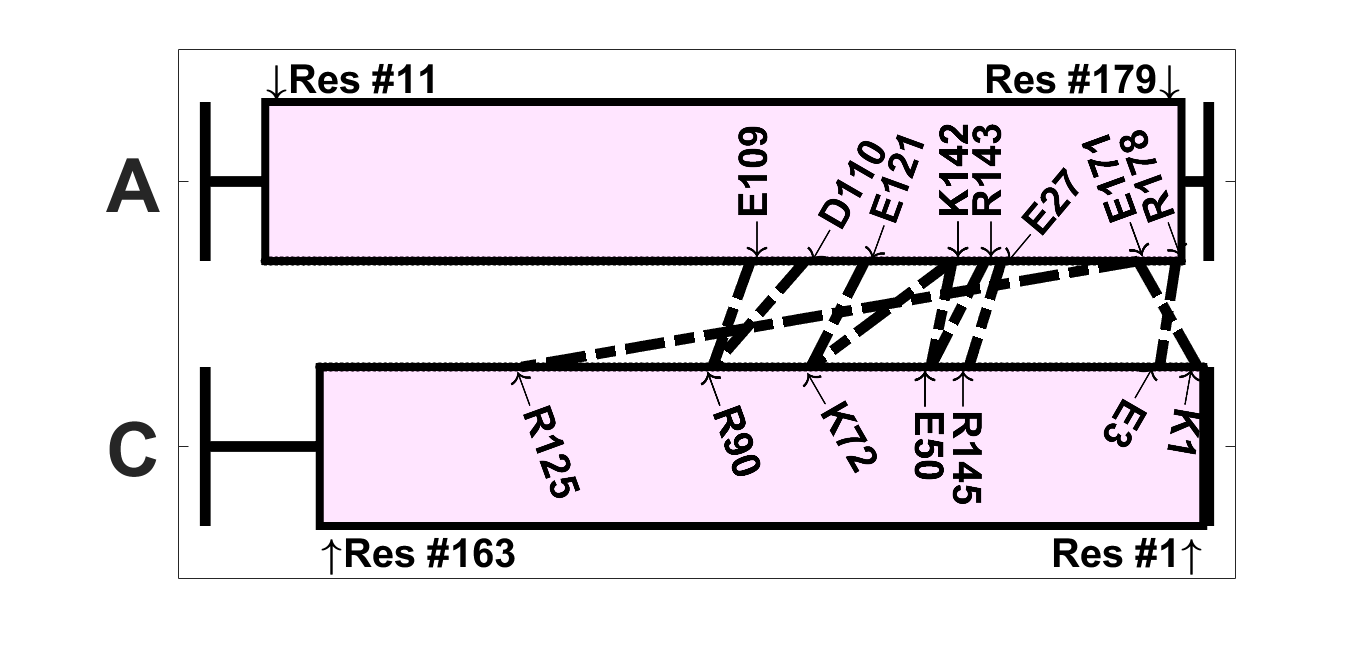}
    \includegraphics[scale=0.18]{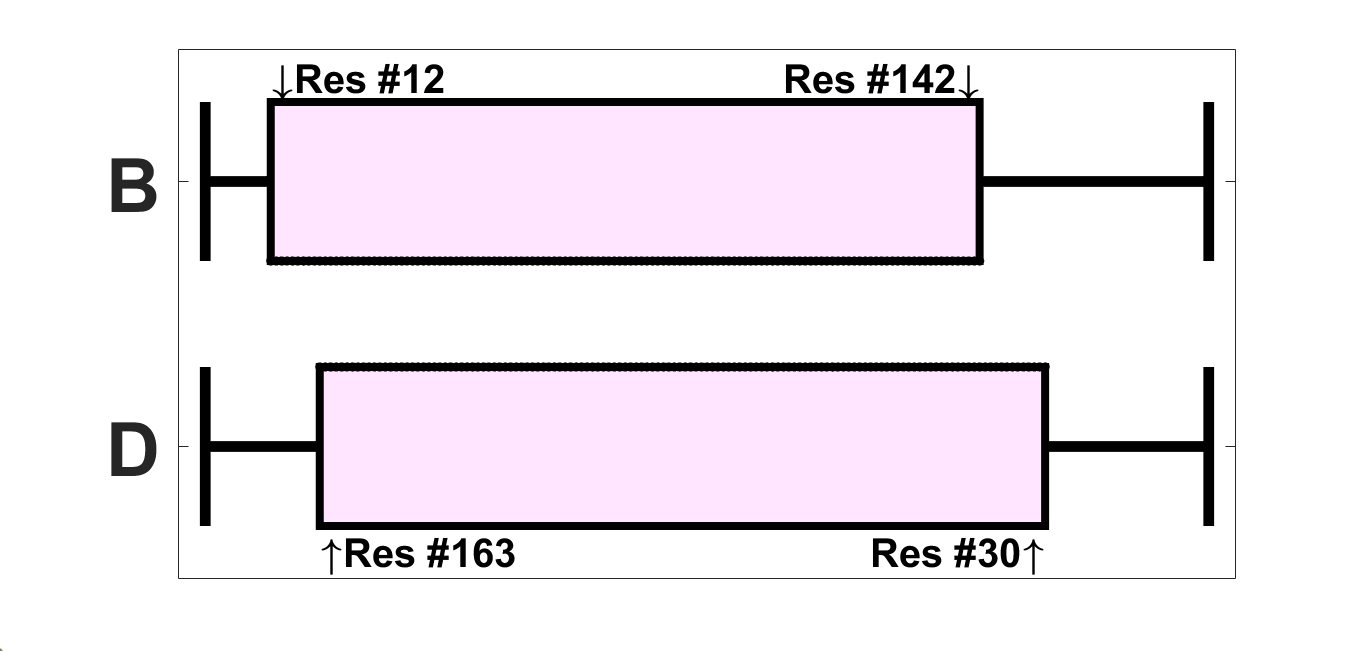}

    420 DMPC molecules\\
    \includegraphics[scale=0.18]{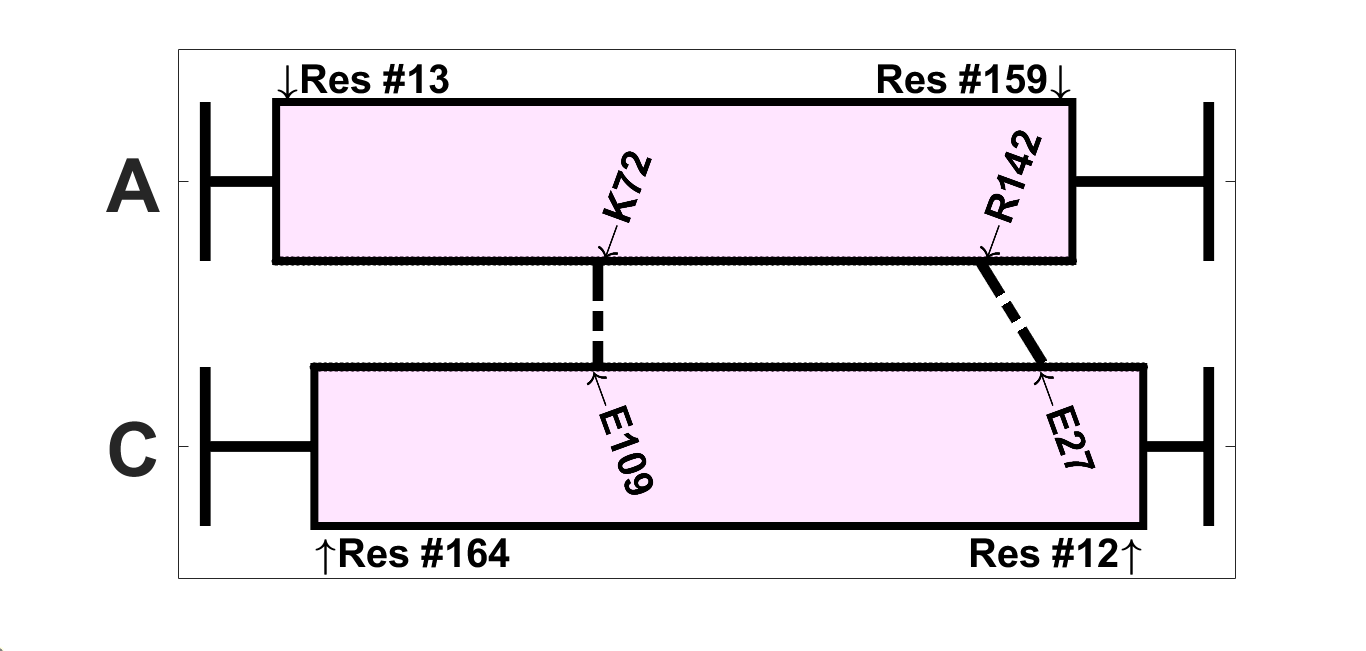}
    \includegraphics[scale=0.18]{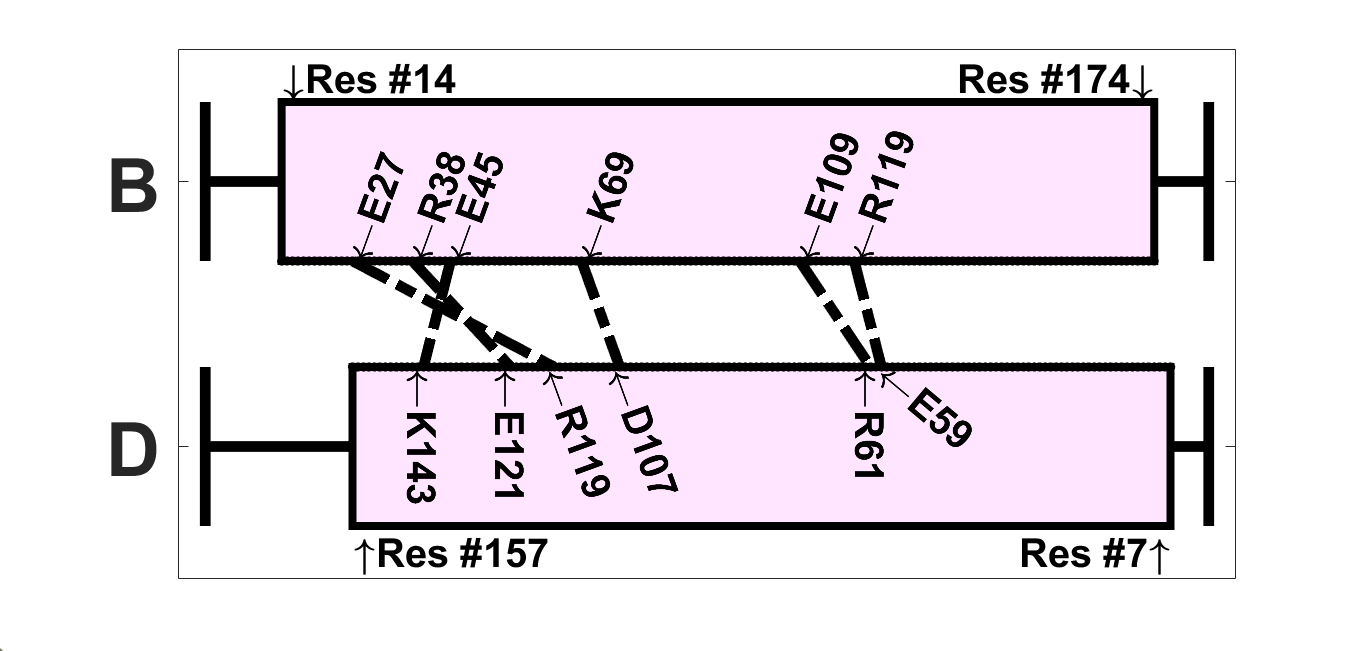}
    \caption{Visualization of the ionic contacts in nanodiscs with antiparallel belt configuration, shown for systems between 240 -- 420 DMPC molecules. The number of contacts is given in Table \ref{tab:contacts}.}
    \label{fig:contacts_2}
\end{figure}

If we inspect the specific amino acids that are involved in the ionic contacts (Figs. \ref{fig:contacts_1} and \ref{fig:contacts_2}), we note that these contacts are not conserved among the different nanodiscs. We notice, however, that a few specific contacts appear in several systems. For instance, the ionic contact between E109 and K75 appears in several antiparallel systems (240-A, 280-A, 320-A, 340-A, 380-A). Also the contact between K72 and E109 appears in 240-A, 340-A, and 420-A.
Interestingly, many amino acids involved in ionic contacts in the antiparallel nanodiscs belong to the group of solvent exposed amino acids in the helix bundle structure of apoE3-NT \cite{sivashanmugam2009unified}. In the helix bundle structure these amino acids are not involved in the interaction between the helices. Examples of these amino acids are K72, E109, R92, E19, R142, E121.
In contrast R119, plays a role in helix bundle stabilization and appears as ionic contacts in several systems (260-P, 280-A, 320-A, 380-A, 420-A);
another example for this behavior is D107, appearing as ionic contact in 320-P, 340-A, and 360-A.

The 420-A system displays the lowest number of total contacts and ionic contacts within the nanodiscs with antiparallel protein configuration. Interestingly, this system also exhibits the much larger backbone RMSD, than the other antiparallel systems (Fig.~\ref{tab:table1}). However, no direct correlation between the number of ionic contacts and the backbone RMSD is found for the remaining systems.
The 380-A system displays the second highest number of total contacts and the highest number of ionic contacts within the nanodiscs with antiparallel protein configuration. Interestingly, this system also shows a relatively high order parameter, as discussed below.

\begin{figure}
    \centering
    260 DMPC molecules\\
    \includegraphics[scale=0.18]{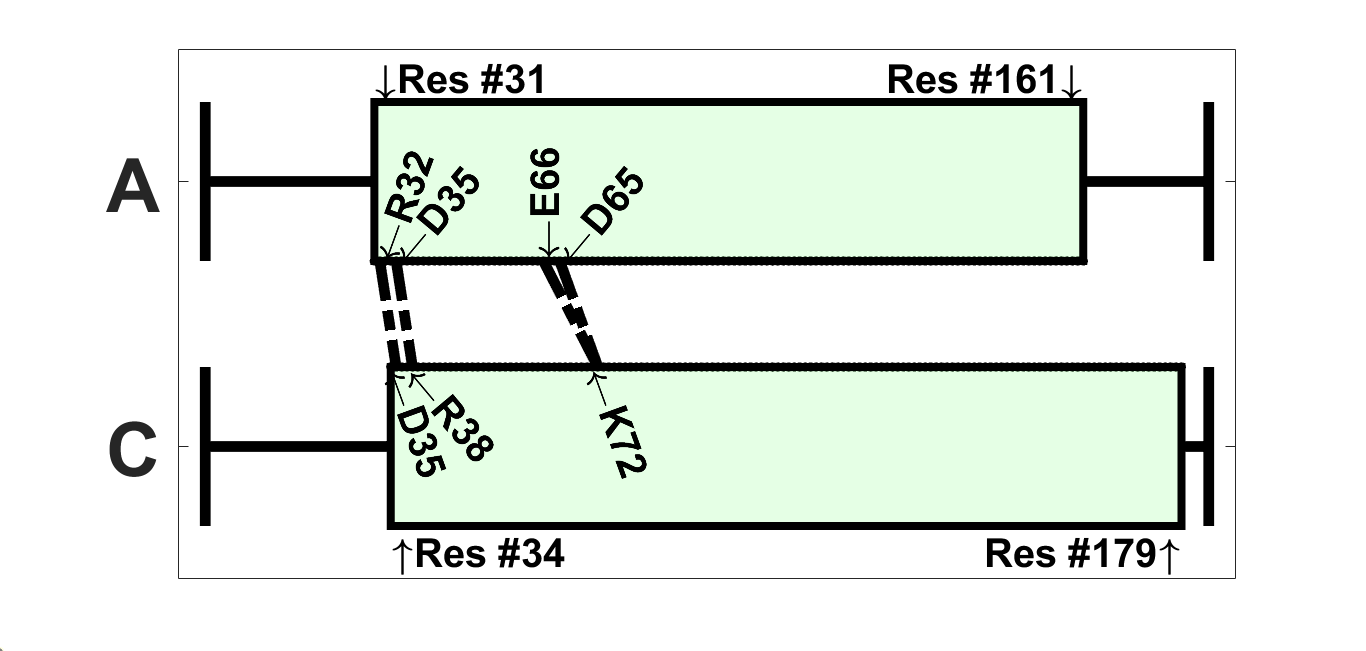}
    \includegraphics[scale=0.18]{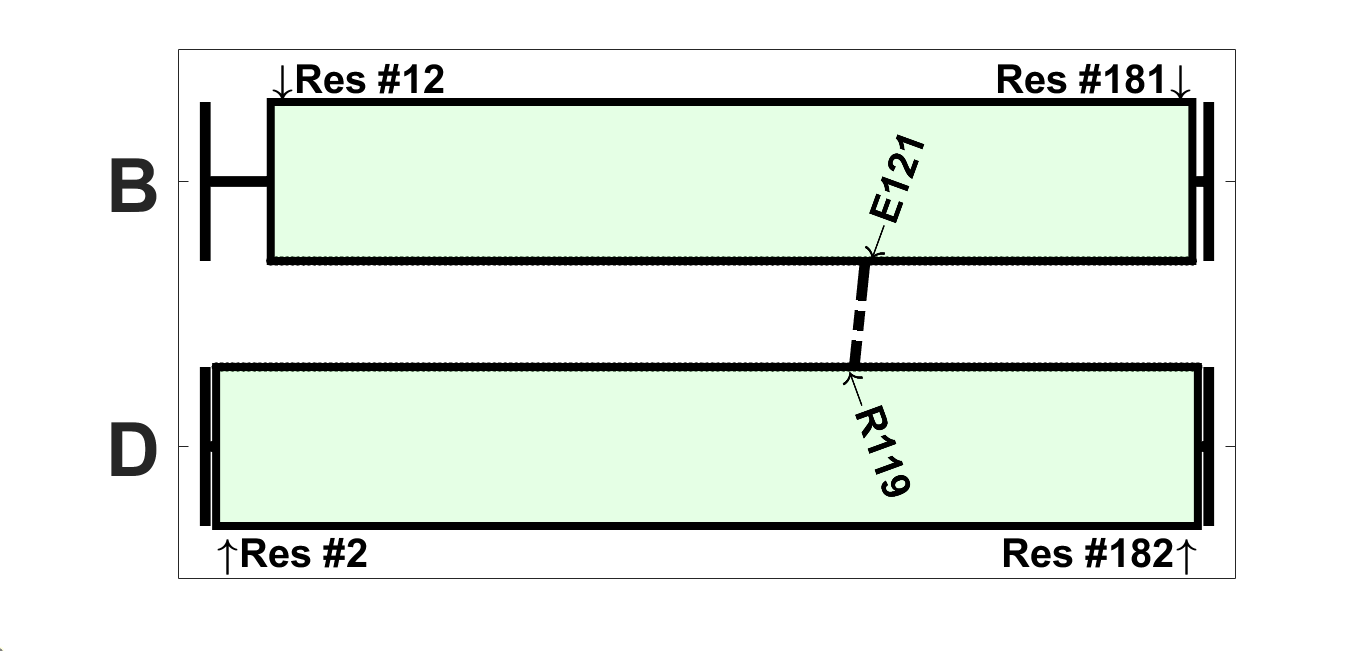}

    280 DMPC molecules\\
    \includegraphics[scale=0.18]{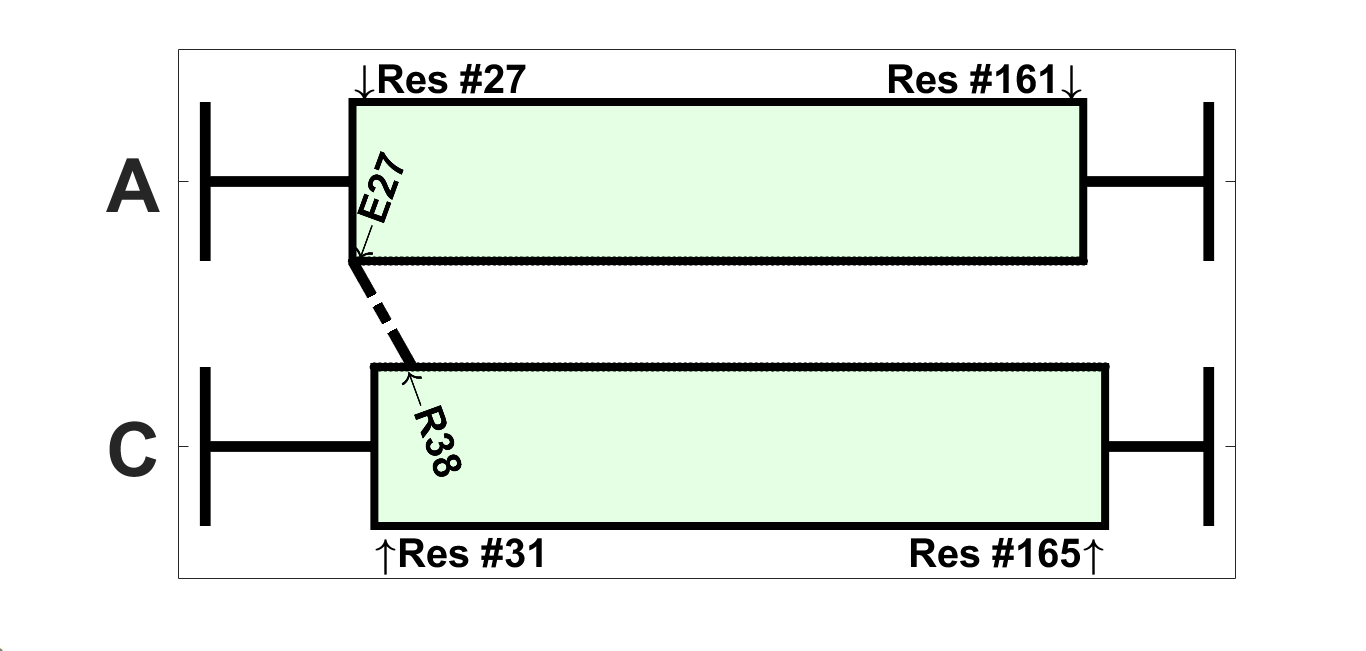}
    \includegraphics[scale=0.18]{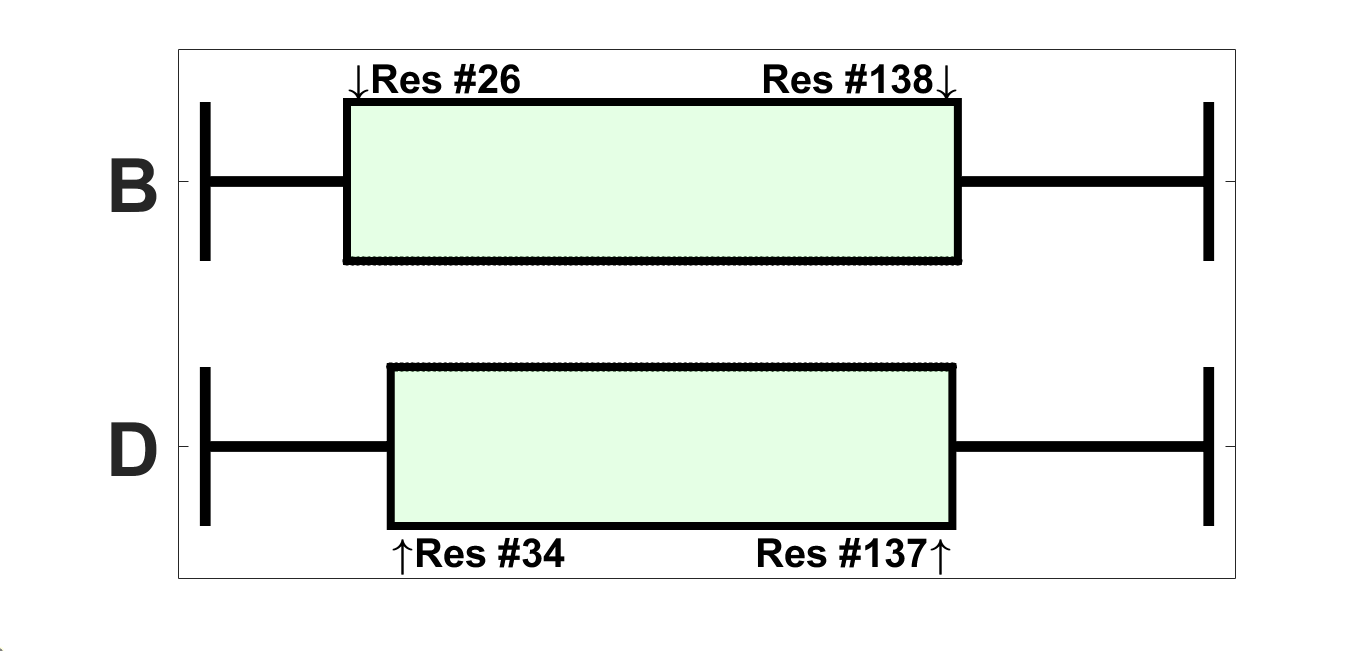}

    300 DMPC molecules\\
    \includegraphics[scale=0.18]{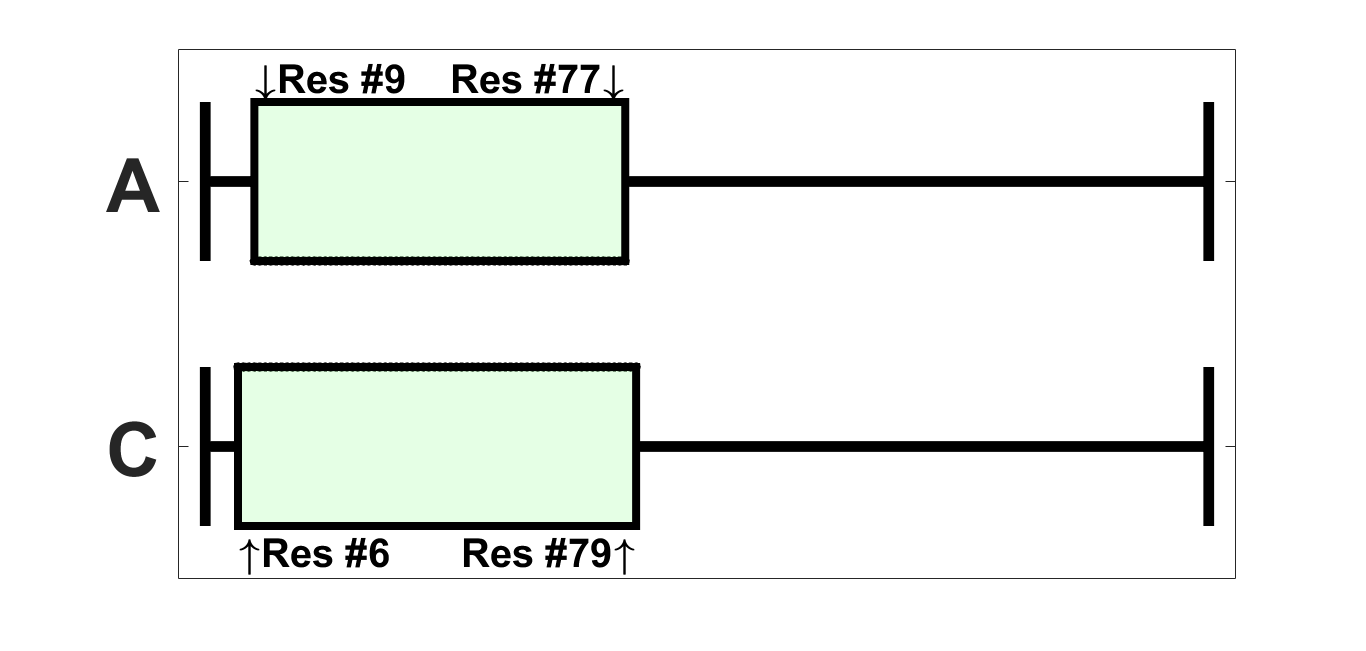}
    \includegraphics[scale=0.18]{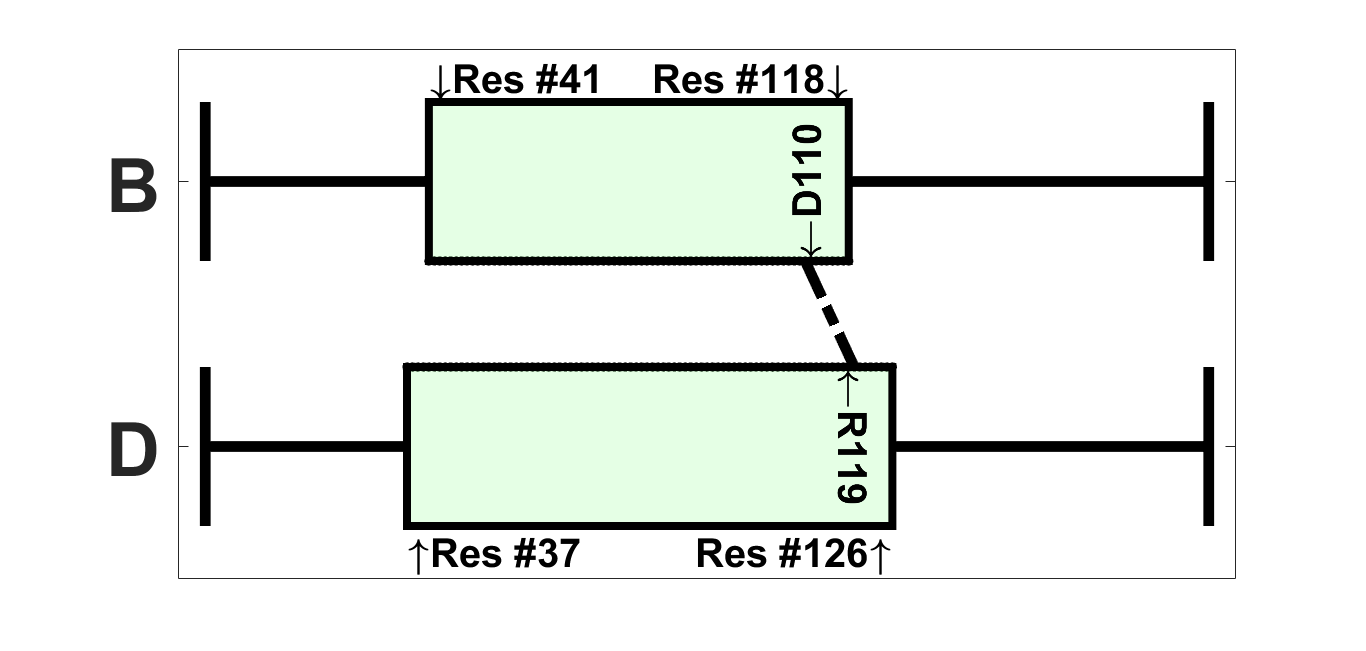}

    320 DMPC molecules\\
    \includegraphics[scale=0.18]{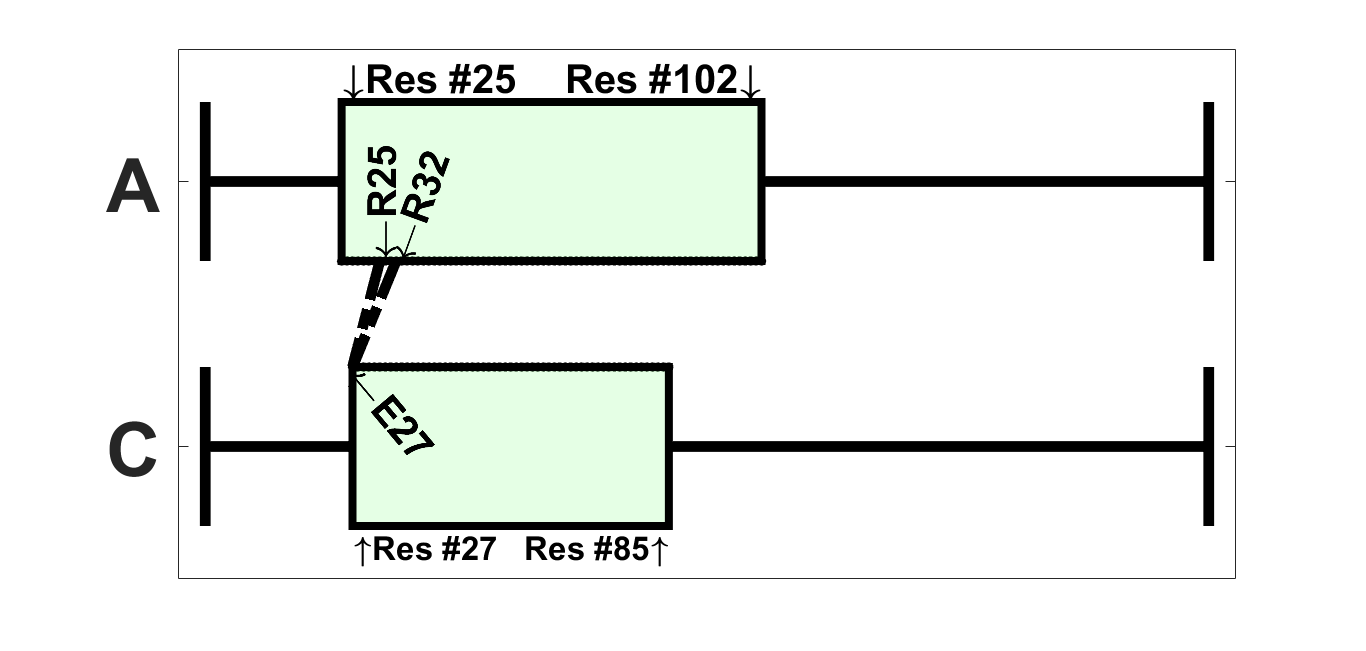}
    \includegraphics[scale=0.18]{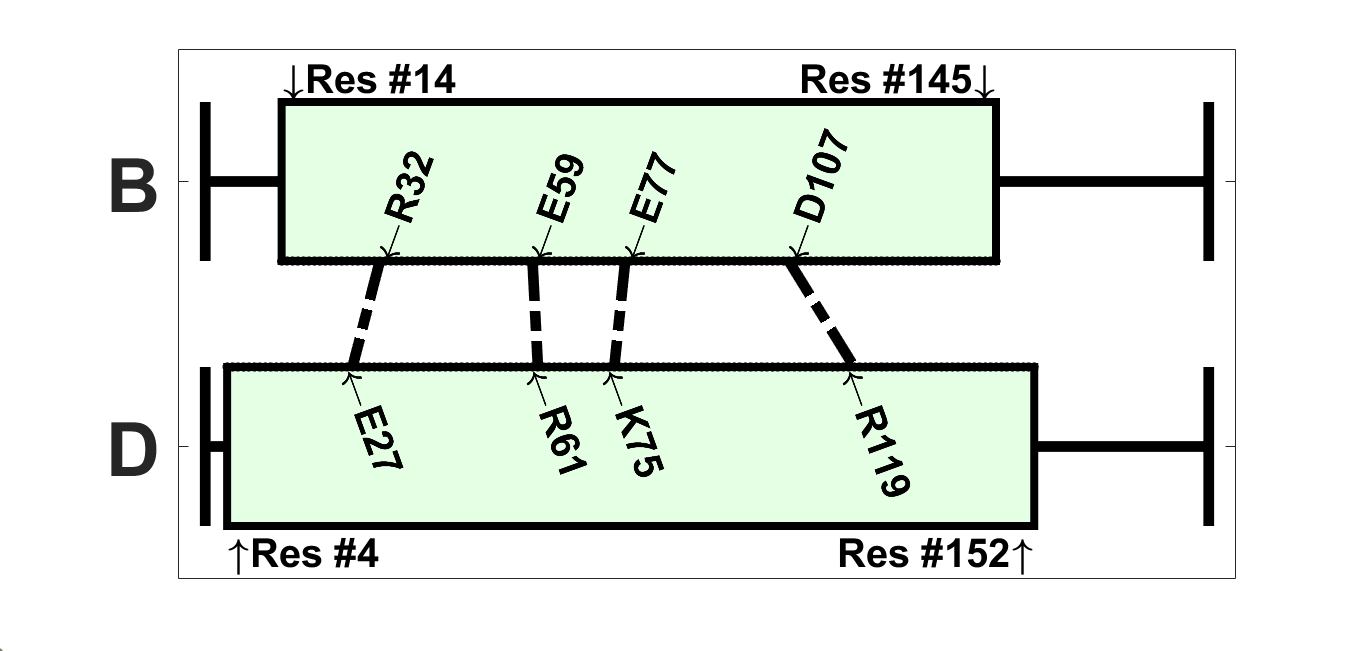}

    340 DMPC molecules\\
    \includegraphics[scale=0.18]{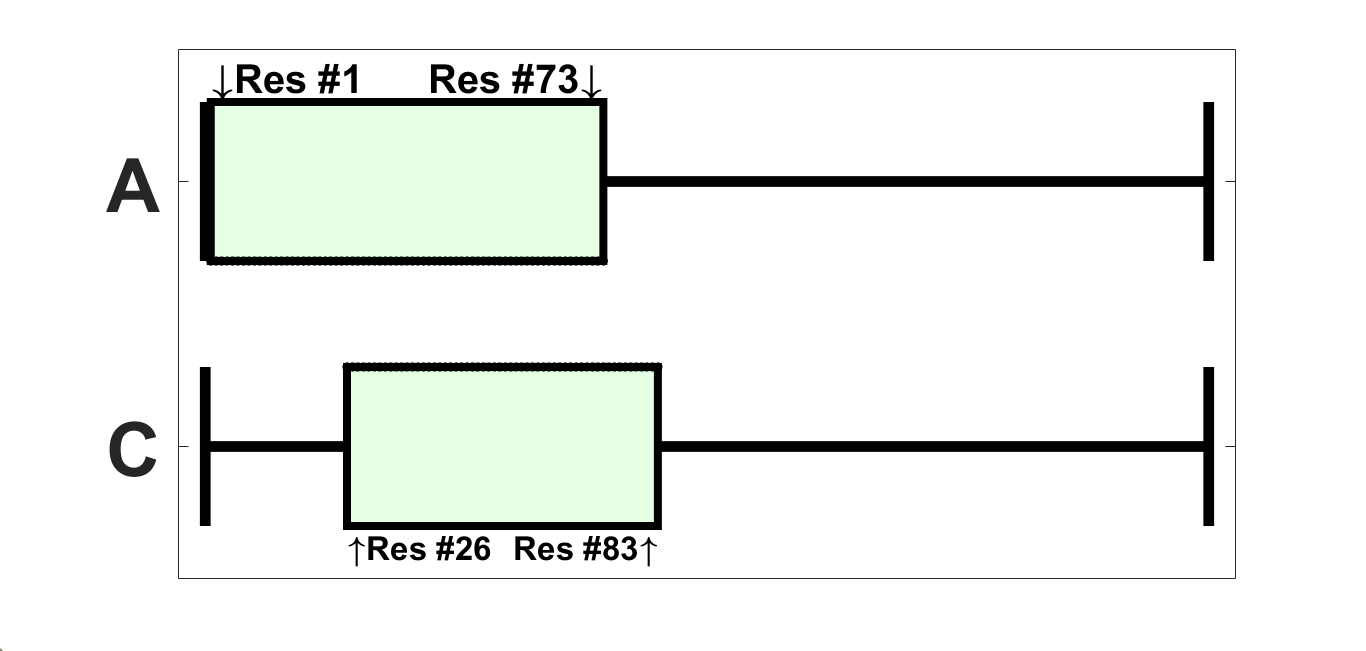}
    \includegraphics[scale=0.18]{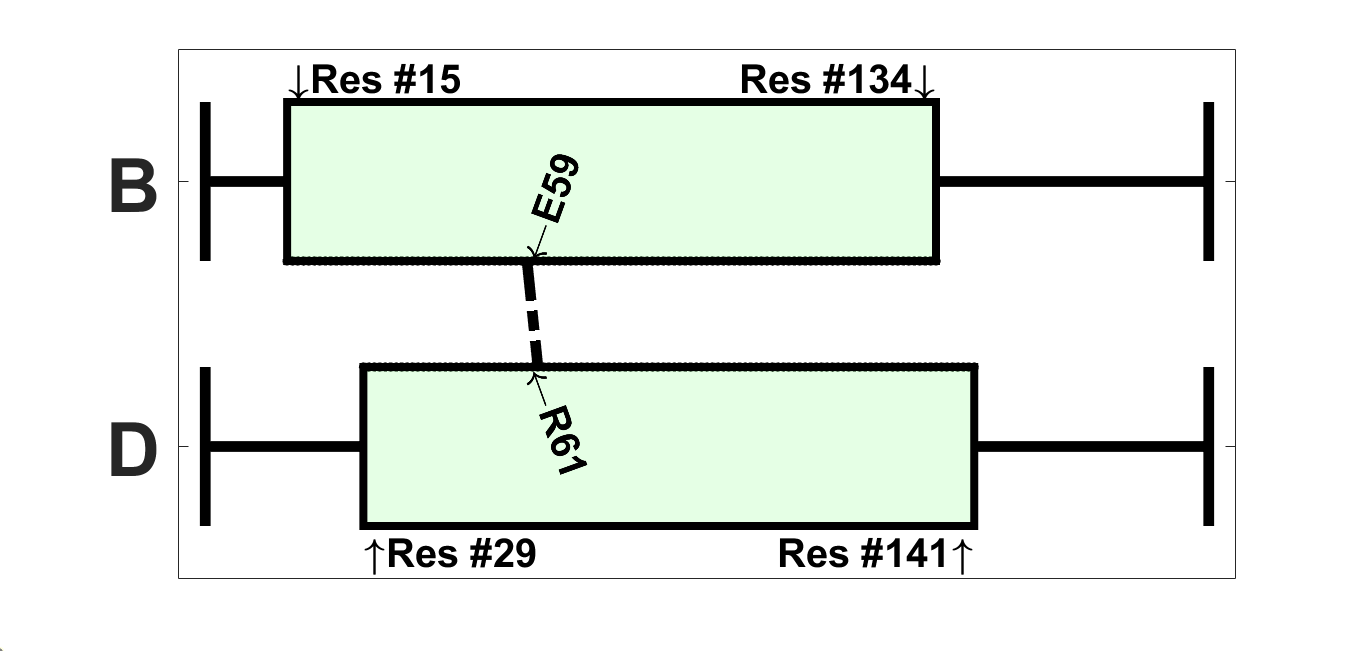}
\end{figure}

\begin{figure}
    \centering
    240 DMPC molecules\\
    \includegraphics[scale=0.18]{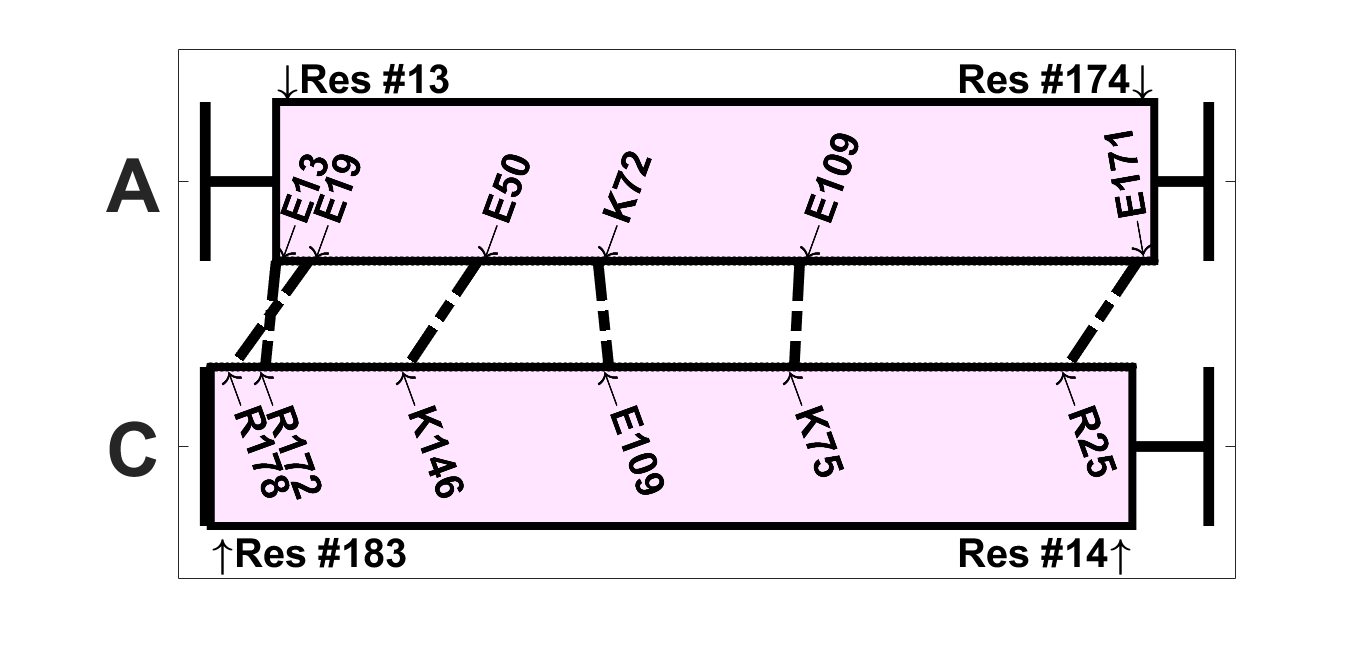}
    \includegraphics[scale=0.18]{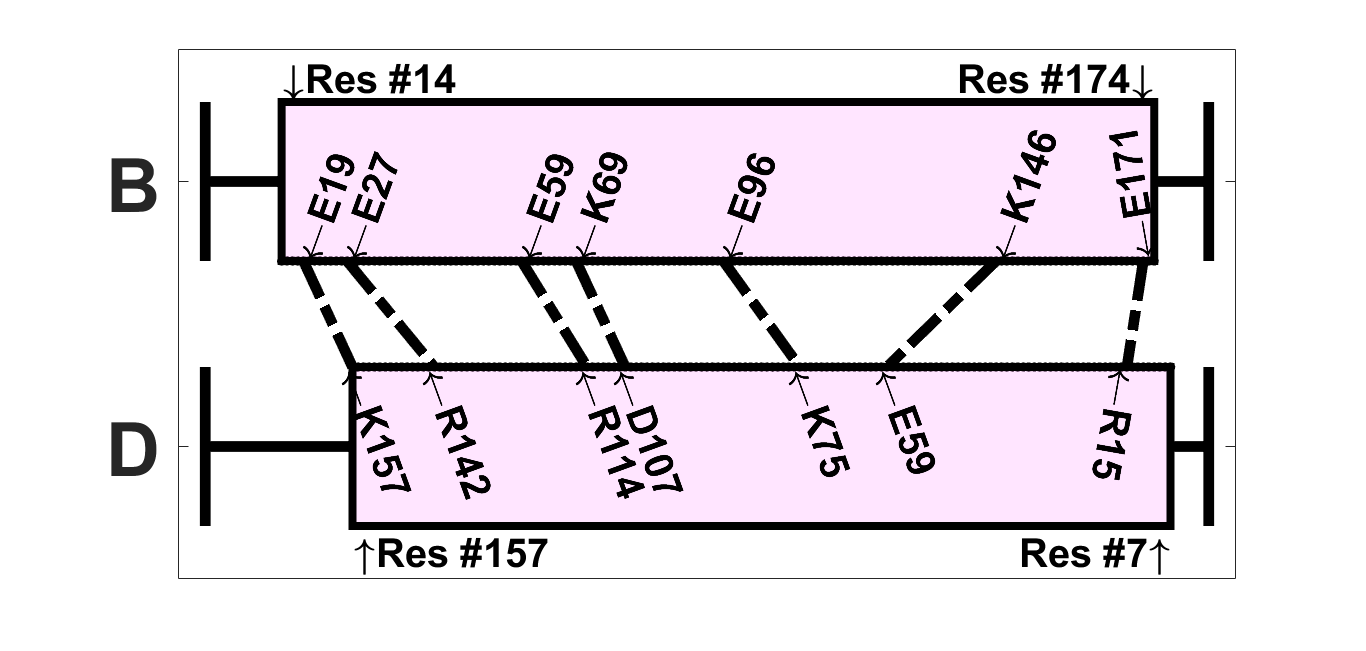}

    260 DMPC molecules\\
    \includegraphics[scale=0.18]{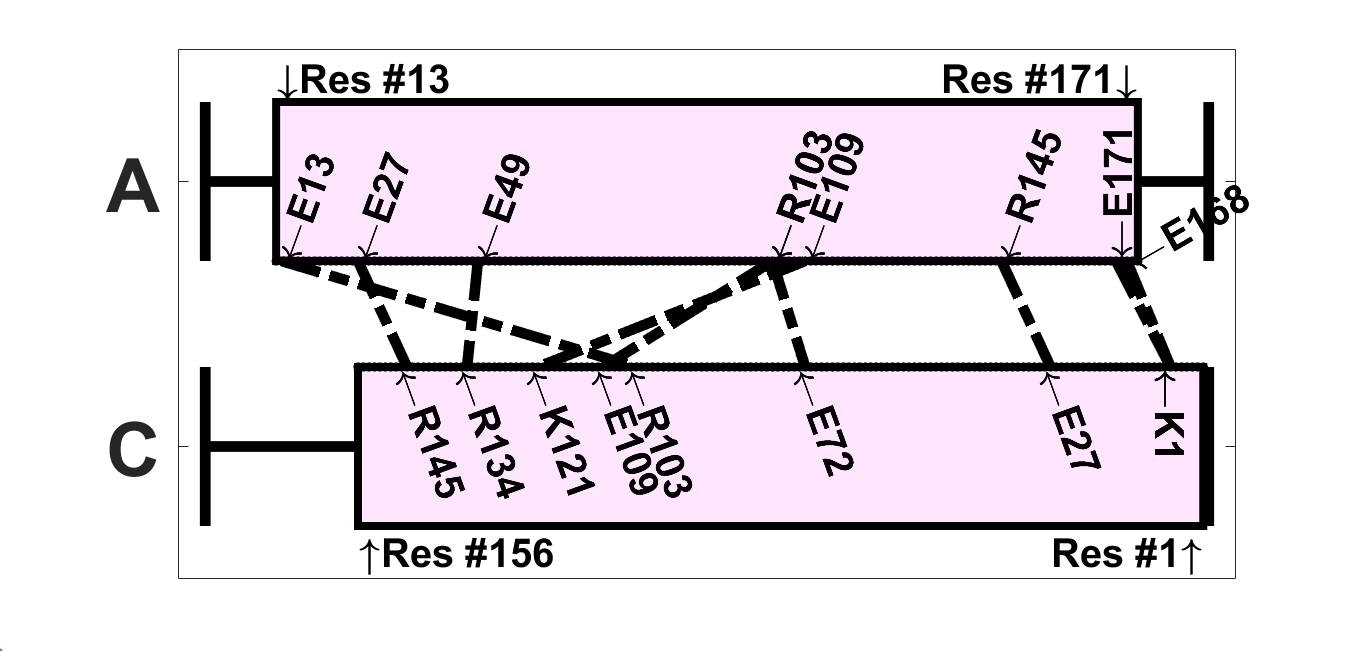}
    \includegraphics[scale=0.18]{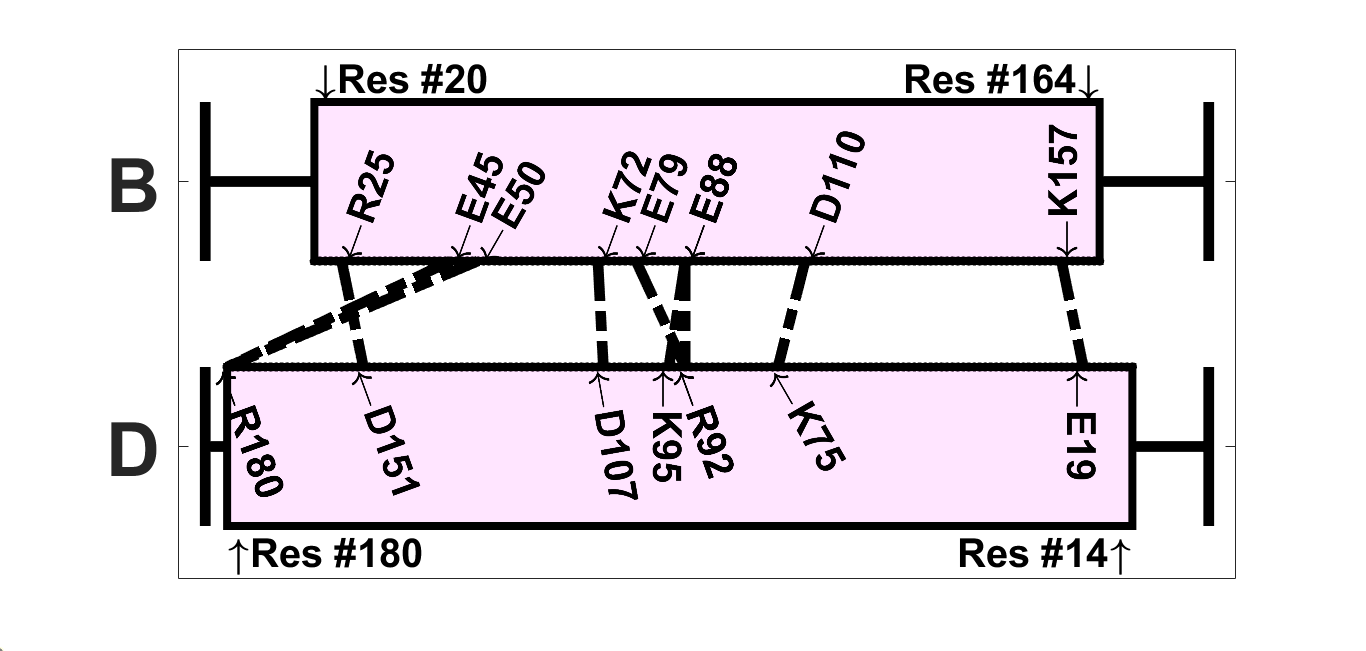}

    280 DMPC molecules\\
    \includegraphics[scale=0.18]{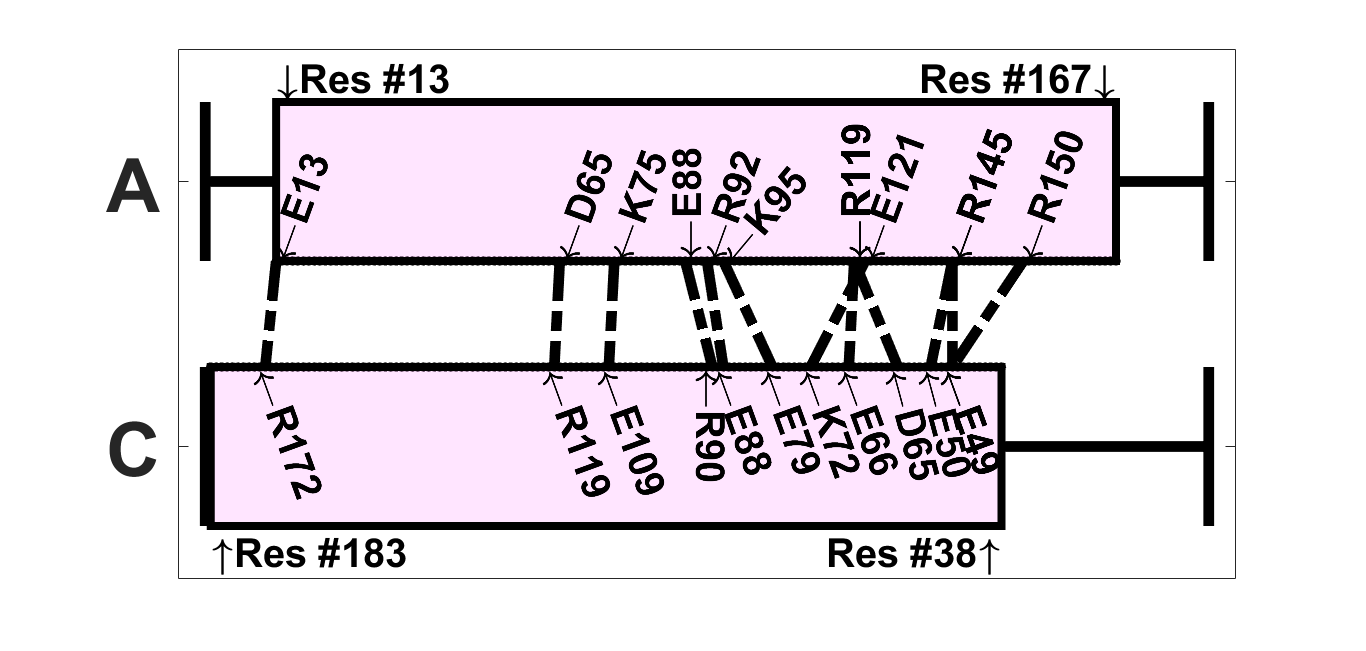}
    \includegraphics[scale=0.18]{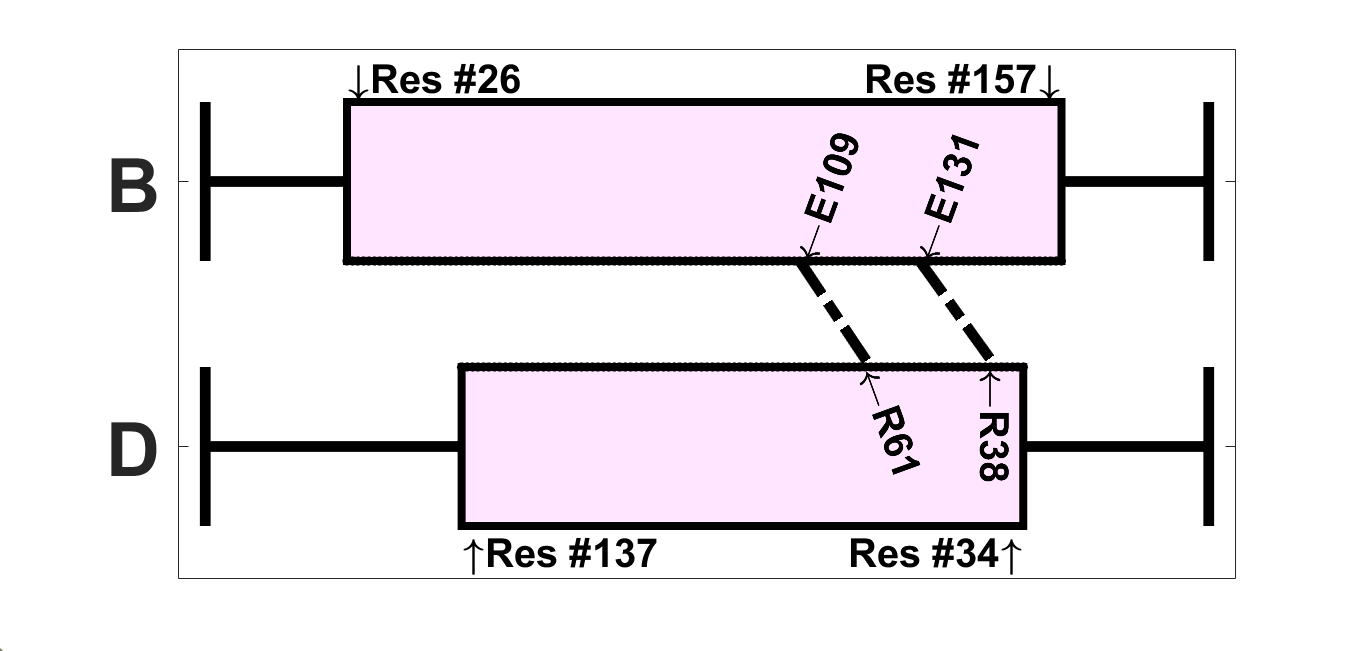}

    300 DMPC molecules\\
    \includegraphics[scale=0.18]{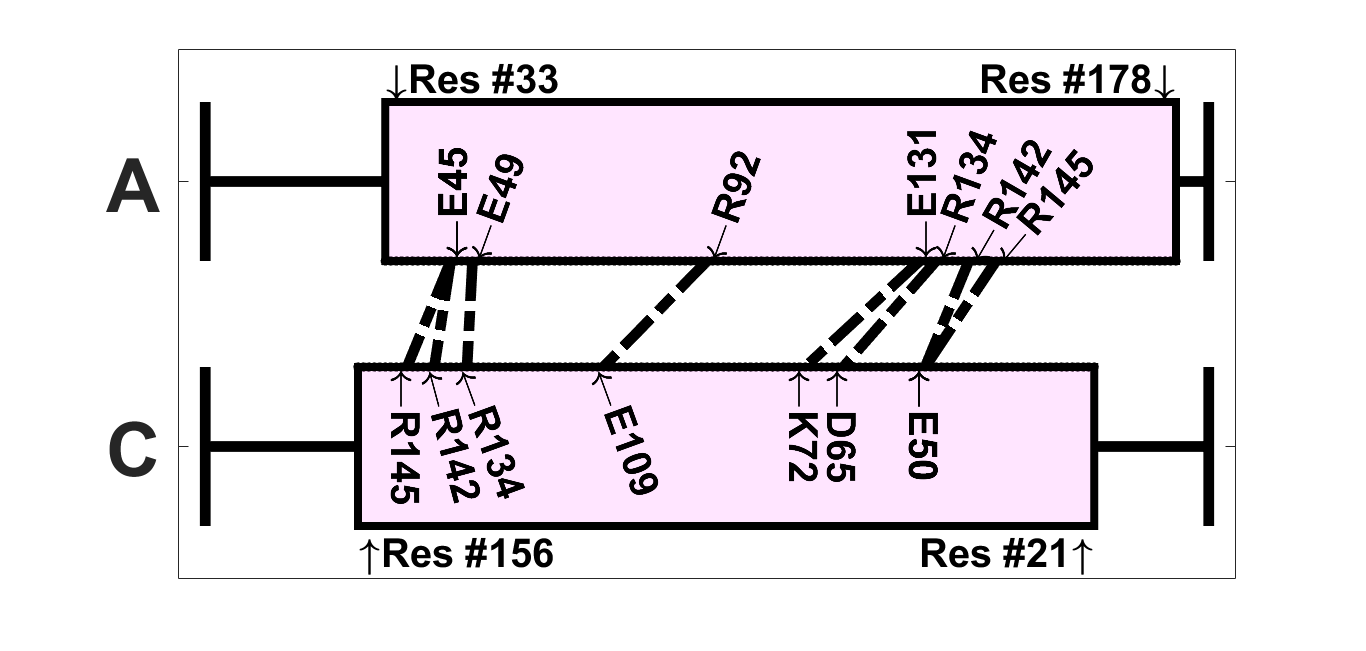}
    \includegraphics[scale=0.18]{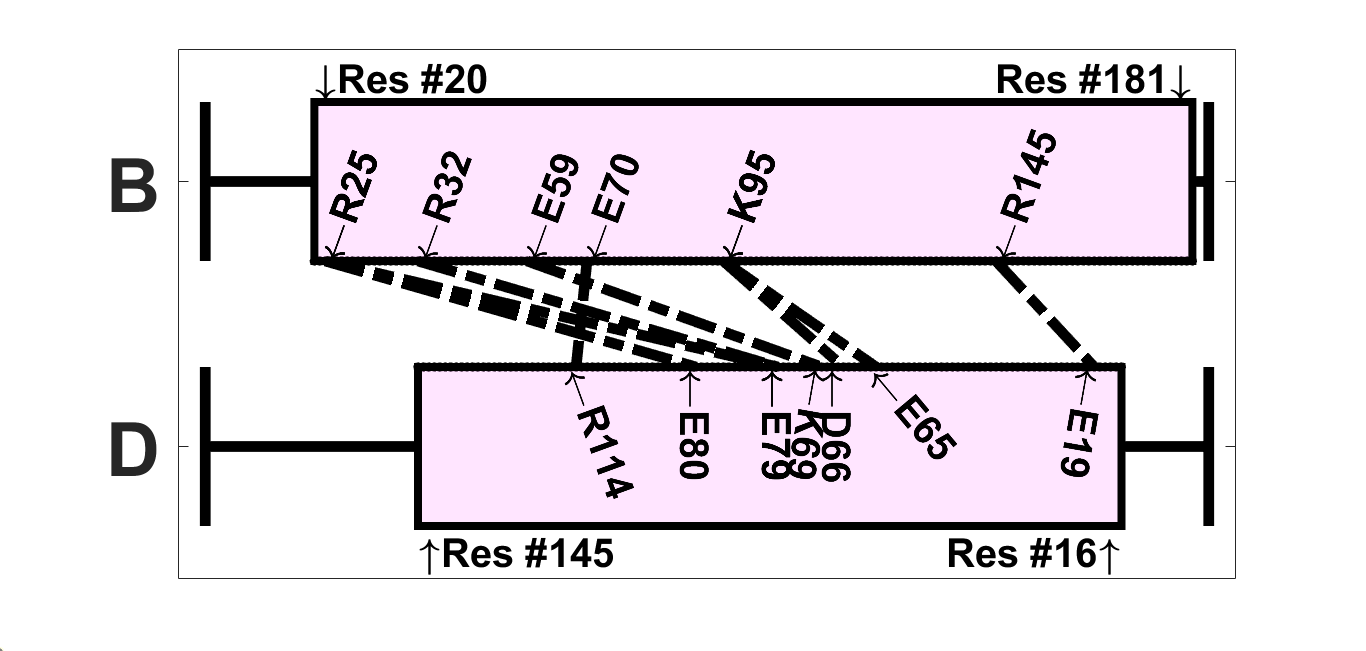}

    320 DMPC molecules\\
    \includegraphics[scale=0.18]{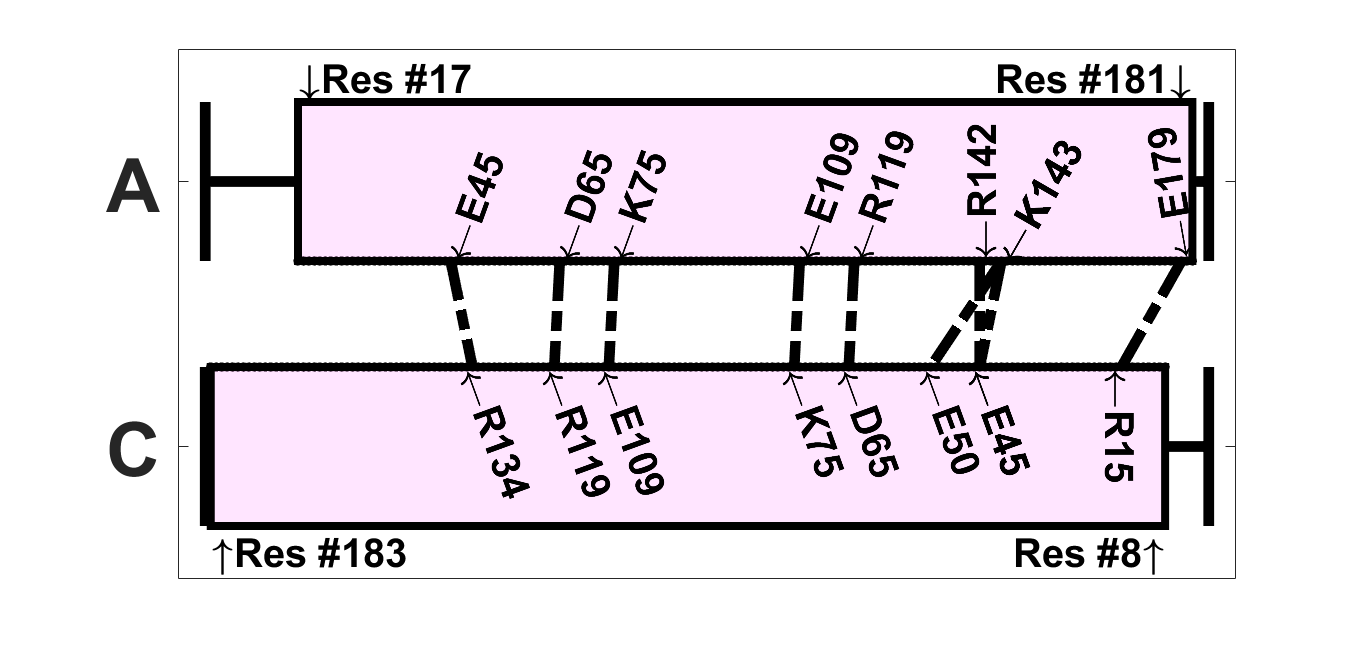}
    \includegraphics[scale=0.18]{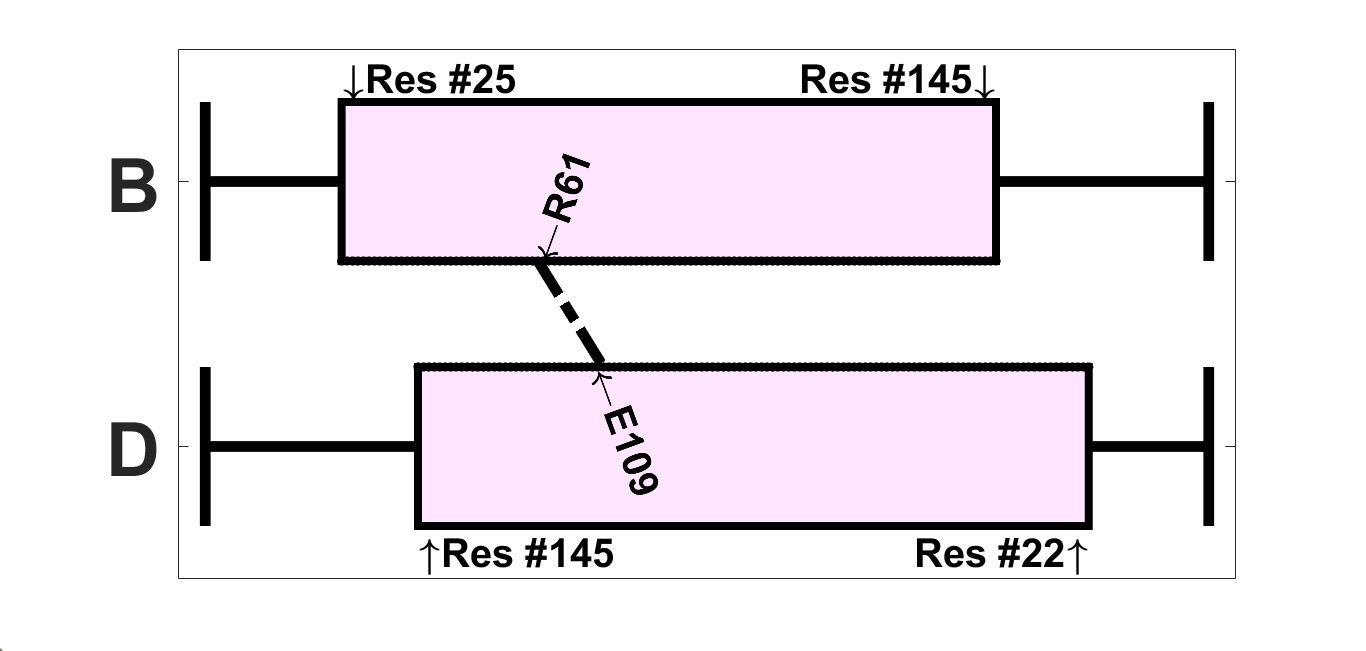}
\end{figure}

\subsubsection*{Order Parameter}
\begin{figure}
    \centering
    \includegraphics[scale=0.26]{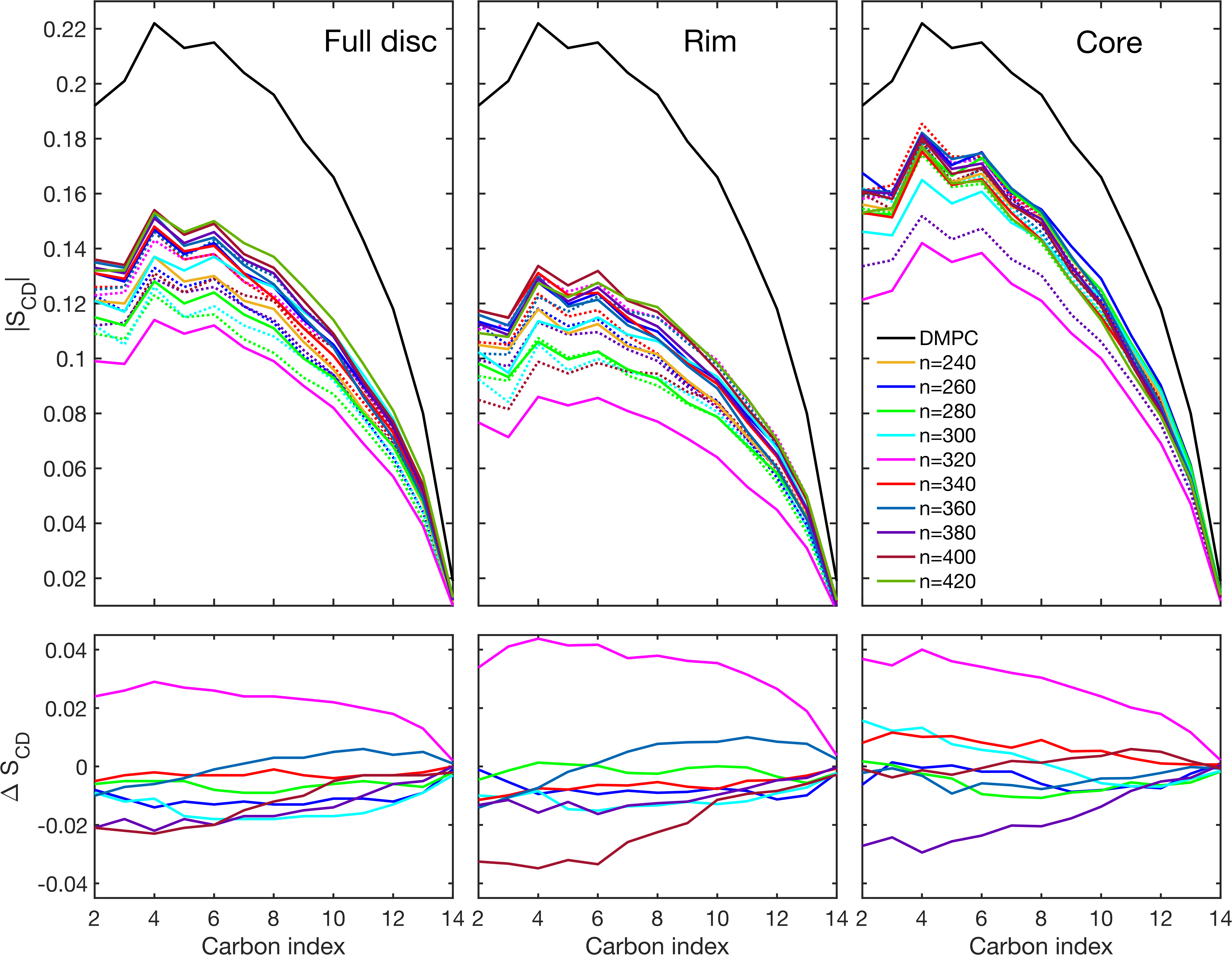}
    \caption{First row: Order parameter ($|S_{CD}|$) as a function of the DMPC carbon index of nanodiscs with different number of lipids ($n$) using all DMPC molecules (full disc), DMPC molecules within 14 nm from the protein (rim), and DMPC molecules in 3 nm from the center of the disc (core). Antiparallel conformation is given in solid lines, parallel conformation is given in dashed lines. DMPC indicates the order parameter for a pure phospholipid bilayer.
    Second row: difference between order parameters of discs with parallel ($\uparrow\uparrow$) and antiparallel ($\uparrow\downarrow$) configuration  $\Delta S_{CD}=|S_{CD}(\uparrow\uparrow)|-|S_{CD}(\uparrow\downarrow)|$. Color code applies to all figures.}
    \label{fig:order_parameter}
\end{figure}

In all nanodiscs, the order parameters measured for all DMPC molecules were lower than for the pure DMPC bilayer (Fig. ~\ref{fig:order_parameter}).
Furthermore, the order parameter of the core is consistently larger than the order parameter of the rim. This has been observed previously in other nanodiscs \cite{bengtsen2020structure}.

Considering the full disc, we note that there seems to be some correlation between the order parameter and the total number of lipids: the larger the number of lipids, the larger the order parameter; this correlation, however, is not strictly obeyed. Most antiparallel systems exhibit larger full-disc order parameters than their parallel counterparts; this can been seen by the negative values of the $\Delta S_{CD}$ value in Fig.~\ref{fig:order_parameter}, lower left panel. However, 320-A exhibits a much lower order parameter than 320-P. Inspecting the structure of 320-A, we see that this is caused by a hydrophilic pore in the center of the nanodisc (Fig.~S38, SI).
Another exception for this behavior are the systems with 360 lipids; here the difference in order parameter is between parallel and antiparallel system is small negative for carbon 1-6 and it becomes positive for carbons 7-14.

The trend of larger order parameters in antiparallel configuration is less obvious, when comparing the core order parameters. While 320-A still has much lower order parameters than 320-P, which is caused by its hydrophilic pore, only 380-A has significantly larger order parameter than its parallel counterpart. As mentioned before, this is the system with highest number of ionic contacts, pointing to a possible relationship between the number of ionic contacts and the order parameter. For a definitive correlation, however we do not have enough data. For the remaining nanodiscs, differences between the order parameter in the parallel and antiparallel configuration are rather small, indicating that further away from the protein belt, the lipid bilayer behaves more similar to a pure DMPC bilayer.

Similar values of the order parameters have been found in DMPC nanodiscs using the proteins MSP1D1 and MSP1E3D1 \cite{stepien2020complexity}. An increased order parameter in the core of the discs was also found. However, in contrast to our study, it was found that the tail of the lipids in the nanodisc showed larger order parameter than in the pure membrane. In our study, order parameters of the pure DMPC lipid membrane were always larger for each carbon atom in the DMPC molecule than in the nanodiscs.


\subsection*{Conclusion}
Our MDS predict apoE3-NT to form stable nanodiscs containing 240-420 DMPC molecules.
Electron microscopy and particle composition of apoE3-NT/DMPC nanodiscs reveal a heterogeneous population with nanodisc particle diameters ranging from 15-20 nm, with ~4 apoE3-NT molecules per nanodisc and with a lipid:protein molar ratio of 150:1 \cite{pitas1980cell,fisher1997bacterial,fisher2000lipid}. Our simulations confirm the approximate diameter and further indicate that the lipid:protein ratio determined from experiments also leads to stable systems in MDS.
Further flotation equilibrium and non-denaturing polyacrylamide gel electrophoresis analyses of the nanodiscs revealed molecular mass ranging from ~440,000 to 760,000 Da. Together, these data suggest that apoE3-NT can accommodate a range of lipids, which supports the current findings of the possibility of stable nanodiscs over a range 240-420 DMPC per nanodisc, corresponding to systems with 298,703 -- 420,730 Da.

The structures obtained from MDS exhibit relatively large RMSD values in the protein backbone positions (0.43 – 0.71 nm) compared to the helix bundle structure (0.042 nm) \cite{sivashanmugam2009unified}.
Thus, our calculations predict nanodiscs to be relatively flexible compared to lipid free protein structure. However, we have to keep in mind that in the structure determination of the helix bundle additional constraints from NMR were used, which could also be the cause of the low RMSD in the lipid free structure.
The high flexibility in our structures does not allow to determine one dominating structural arrangement between protein and DMPC molecules. It rather lets us conclude that there might be a wide range of structural arrangements that lead to stable nanodiscs. The limited sampling and the assumption in the generation of the initial structures, i.e. parallel and antiparallel protein configuration, allow us to only give a glimpse of the possibilities of structures that might exist.
An overall assessment of the ensemble of possible nanodisc structure would require more sophisticated sampling methods \cite{pourmousa2018molecular}.
However, our study could serves a possible starting point for such further investigations.

Despite the limitations of our set-up, there are some conclusions that can be drawn from this study regarding apoE3-NT nanodisc behavior. First of all, our simulations indicate a double-belt like structure of the apoE3-NT protein.
The protein configuration (parallel vs. antiparallel) of apoE3-NT chains appears to affect the stability and rigidity of the nanodiscs.
It appears that regarding several parameters associated with nanodisc stability, that antiparallel protein configuration leads to more stable and more rigid systems than in the case of parallel protein configuration.
Particularly, the number of ionic interactions between amino acids of different chains is much larger in the antiparallel configuration.
Interestingly, most of the polar side chain--solvent interactions in the lipid-free state are replaced by polar side chain--polar side chain interactions in the lipid-associated state.
However, we find also ionic amino acids that are involved in protein-protein interactions in both, the helix bundle and the nanodiscs.

Furthermore, the results for the order parameter suggest that the DMPC molecules in antiparallel protein configurations are more ordered compared to parallel configurations.
The generally larger order parameter in antiparallel systems, together with their generally larger number of ionic contacts suggest a crucial contribution of the ionic contacts to nanodisc stability, which has been found in other nanodisc systems \cite{mei2011crystal,xu2023reconfigurable}.

In the future, experimental observation of the preferred protein arrangement could be obtained by solution NMR spectroscopy \cite{puthenveetil2017nanodiscs,gunsel2021lipid}. Enhanced sampling methods \cite{pourmousa2018molecular} could provide a more complete assessment of the possibilities of structures.


\subsection*{Acknowledgments}
Research reported in this publication was supported by the National Institute of General Medical Sciences of the National Institutes of Health (NIH) under award numbers R25GM071638, G238620100 and R15GM126524, 1 R16GM149410-01, BUILD Small Equipment and Computers Grant UL1GM11897G-02, TL4GM118980, RL5GM118978, GM 105561, NSF LSAMP (HRD 182649), and the Carl E. Riley Endowed STEM Award. The authors thank Justin Lemkul for his well-designed GROMACS tutorials and Dmitri Rozmanov for help troubleshooting bio.b-gen.
The content is solely the responsibility of the authors and does not necessarily represent the official views of the NIH. We also acknowledge technical support from the Division of Information Technology of CSULB.

\section*{Supplementary Material}
Molecular visualizations of the final structures for CG and AA simulations, Tables with SASA, radius of gyration, and Protein contact maps are given in the SI.

 \bibliographystyle{elsarticle-num} 
 \bibliography{references}





\end{document}